\documentclass[11pt]{article}

\usepackage[margin=1.4in]{geometry}

\usepackage{diagbox}

\usepackage{pgfplots}
\usetikzlibrary{trees}
\usetikzlibrary{calc}

\pgfplotsset{compat=newest}

\usepackage{amsmath,amssymb,amsthm,enumerate}

\usepackage{tikz}
\newtheorem{Theorem}{Theorem}

\newtheorem{Proposition}{Proposition}
\newtheorem{Definition}{Definition}
\newtheorem{Lemma}{Lemma}

\usepackage{graphicx}
\usepackage{float}

\usepackage{endnotes}

\newcommand{\gi}{G \vert I}        %% game G subject to information function I(x)
\newcommand{\mix}[1]{M_{I(#1)}}

\usepackage[authoryear]{natbib}		% removed
\bibpunct{(}{)}{;}{a}{,}{,}			% removed

\begin{document}

\begin{center}

\medskip

%{\Large Strategic games with purchase of}
%\medskip
%
%{\Large randomly supplied information: Oracle Games}
%\medskip

{\Large Simultaneous games with purchase of}

\medskip

{\Large randomly supplied perfect information: Oracle Games}

\bigskip
\bigskip

Matthew J. Young and Andrew Belmonte\footnote{Email: belmonte@psu.edu}

\medskip

{\it Department of Mathematics, The Pennsylvania State University,\\ University Park, PA 16802, USA}

\medskip

% \today

(February 19, 2020)

\end{center}

\begin{abstract}

We study the role of costly information in non-cooperative two-player games when an extrinsic third party information broker is introduced asymmetrically, allowing one player to obtain information about the other player's action. This broker or ``oracle" is defined by a probability of response, supplying correct information randomly; the informed player can pay more for a higher probability of response. We determine the necessary and sufficient conditions for strategy profiles to be equilibria, in terms of how both players change their strategies in response to the existence of the oracle, as determined by its cost of information function.
% We find that games with a unique equilibrium  will still have a unique equilibrium after an Oracle is added.  Additionally, this equilibrium will change based on the cost of information from the oracle: 
For mixed strategy equilibria, there is a continuous change as information becomes cheaper, with clear transitions occuring at critical {\it nodes} at which pure strategies become dominated (or undominated). These nodes separate distinct responses to the information for sale, alternating between regions where the paying player increases the amount of information purchased, and regions where the other player 
moves away from riskier strategies, in favor of safer bets that minimize losses. We derive conditions for these responses by defining a value of information. 

\end{abstract}

%Keywords: Costly acquisition of information; Asymmetric information

\section{Introduction}

All decisions are made in the presence of information, and sometimes in spite of its absence.  In the classic theory of simultaneous non-cooperative games, information appears in two different ways: while players typically have complete and common knowledge of the rules, strategy choices available, and payoffs in the game, they usually have no information on what actual strategy is about to be played by their opponent. Yet in the real world, people often look to obtain an edge in competing with adversaries by looking for any hint as to what the other player will do; such information could be highly valuable. In fact, there is experimental evidence that human subjects will pay to know what strategy is being played against them, even when that information has no impact on their strategy choice \citep{eliaz2010}. Subjects will also sometimes attempt to deceive an opponent about their own intended strategy \citep{Mcdonald1996}. Beyond formalized game situations, competitors in business and biology are often willing to expend time, capital, or valuable energy resources in order to obtain information to potentially tip the scales beyond the uncertainty of strategy choice \citep{morris2002,asahina2008,Gabaix2006}.
The scenario we consider here for two-player games allows us to define a {\it value of information}, and to study the role it plays among other choices made.

In general, the lack of information in game theory falls into one of two main categories: 
{\it incomplete information}, where players are uncertain about some of the rules of the game or the payoffs that will result from outcomes, and 
{\it imperfect information}, where players lack information about the current state of the game, such as decisions made by other players or random events that have occurred in secret.  Under certain assumptions a game with incomplete information will be equivalent to a Bayesian game with complete but imperfect information \citep{Harsanyi1967}. However, generally the two types of information and their motivations from real world scenarios remain distinct.

In games of incomplete information, the role of information and the willingness of players to purchase it has been studied in a number of contexts, such as 
Beauty Contest games \citep{myatt2012, hellwig2009, rigos2018},
election models \citep{martinelli2007},
Cournot games \citep{myatt2015},
the Battle of the Sexes \citep{Hu2018}, 
and investment games in the presence of noisy signals determined by some underlying state \citep{yang2015, szkup2015}.

Here we focus on games with imperfect information, where one player has the ability to acquire information about the second player's strategy. 
Classic game theory considering one-shot simultaneous games has complete but imperfect information, since neither player knows the strategy of the other player until they have both played out their choices. The value given to advance information on an opponents choice will clearly depend of the arrangement of payoffs and strategies in the game. 
In treating this situation, it is necessary to consider both the actions of the first player in acquiring this information and responding to it, as well as the second player's actions taking into account the possibility of their strategy being revealed.  Many studies have investigated these issues, for instance iterated games with information about past strategy choices restricted by %behind 
some cost \citep{ben-porath2003, flesch2009, miklos2013}.  
Other studies include the difference between Cournot or Betrand duopoly games and Stackelberg games in terms of information sharing \citep{ruiz2017,sakai1986}, games with random information leaked revealing possible changes in strategy \citep{halpern2018}, and iterated Prisoner's Dilemma with network effects \citep{antonioni2014}.
%%%%%%%% Spy games
\citet{solan2004} study a modification to two-player games in which one player can purchase a noisy signal correlated with the opponent's action, but has some probability of signalling a different action and misleading the player.  Higher payments increase the reliability of the signal, depending on the cost function associated with the signalling device.

We introduce a formalism into standard two player games for the purchase of information about the realization of strategy choices, and study its effect on pure and mixed strategy equilibria.  
We do not treat the theoretical issues surrounding the meaning of mixed strategy equilibria in games (see e.g.~\cite{reny04}), focusing instead of the implementation of such equilibria in individual one-shot games (see for instance the analysis of soccer penalty kicks by \cite{chiappori02}). In a given game, after a player has chosen to play a pure or mixed strategy, the latter choice will involve a second step in which an independent process is required to select at random one of the pure strategies to be played, according to the probabilities of the strategy. It is the information about this final selection that we are concerned with here.

We replace the standard approach to partial information, in which players pay for the increased accuracy of noisy signals, with a partial response approach in which completely accurate information is purchased but not always received. 
Thus any information supplied is always correct, while when information is not supplied, that too is well known to the purchaser.   
The process is intrinsically asymmetric: only one of the players can pay for the probability of learning the particular strategy which will be realized on the other player's side. This other player, meanwhile, knows only that this information may be revealed, and adjusts accordingly. 

%How does the availability of this information affect the equilibria of the game? 

%only sometimes received, and players pay for a higher probability of receiving it.

This paper is organized as follows. We first define an extrinsic third player into a classic two player game: an ``oracle" who can be paid for a chance to reveal information about one of the players to the other. After exploring the consequences of such an oracle in representative examples, we define the properties of these games, and prove results on how any mixed Nash equilibria will be modified by the cost function of the oracle. 
We also discuss the apparent impossibility of including a second, symmetric oracle, and discuss further directions for the development of these ideas.

%%%%%%%%
\section{Preliminary Considerations}

We begin by considering a standard normal form game $G$ with exactly one mixed strategy Nash equilibrium; as we will show later, the modifications we propose here do not affect the pure strategy Nash equilibria, so games with only pure strategy equilibria will be unchanged.  We briefly discuss games with multiple mixed equilibria in Section \ref{multipleequilibria}.

We define an {\it oracle} to be an external agent to $G$ who knows and can potentially reveal information about each player's actual choice of strategy, before these choices are ``played" and payoffs resolve. The oracle is defined to have an associated ``oracle function" $I(x)$ which determines its probability of response as a function of the amount it is paid.  When paid $x$, the oracle either reveals completely accurate information about a player's strategy choice with probability $I(x)$, or remains silent and gives no information with probability $1-I(x)$. In this way, the oracle allows for partial purchase of information about a player's choice without introducing anything other than factual information (i.e.~the oracle either tells the truth or says nothing).

In principle, $I$:$~[0,+\infty) \rightarrow [0,1]$, however the domain of $I$ may be effectively bounded due to a rational player not paying beyond some fixed amount $x_m$, determined for instance by the largest variation in payoffs in the game, $x_m < P_{max}-P_{min}$. Note also that $x=0$ is included, which represents not paying the oracle at all.

%\newpage

\subsection{Motivating Example 1}

To illustrate our approach, we first consider the following two-player normal form game 

\begin{table}[H]
\renewcommand{\arraystretch}{1.1}
\Large
%\Large (used to make the table larger than it needed to be
\centering
  \begin{tabular}[h]{r|c|c|c|}
 \multicolumn{1}{c}{\diagbox[height=1.9em]{\normalsize 1}{\normalsize
    ~2}} & \multicolumn{1}{c}{$B_1$} & \multicolumn{1}{c}{$B_2$}  \\\cline{2-3}
    $A_1$   & $1,-1$ & $0,0$ \\\cline{2-3}
    $A_2$ &  $0,0$ &  $2,-2$ \\ \cline{2-3}
  \end{tabular}
\end{table}

\noindent
which is a matching pennies (anticoordination) game with scaled payoffs.
This game has only one Nash equilibrium, for which A and B each play the mixed strategy $(\frac{2}{3},\frac{1}{3})$.

%%===================================
% tree layout
\tikzstyle{level 1}=[level distance=20mm, sibling distance = 40mm]
\tikzstyle{level 2}=[level distance=20mm, sibling distance = 25mm]
\tikzstyle{level 3}=[level distance=20mm, sibling distance = 13mm]
% node styles
\tikzstyle{player} = [circle, draw, inner sep=1.2, fill=black]
\tikzstyle{end} = [circle, draw, inner sep=1.2, fill=black ]

\newcommand{\payoff}[4][below]{\node[#1] at (#2) {$(#3,#4)$};}
%%%%%
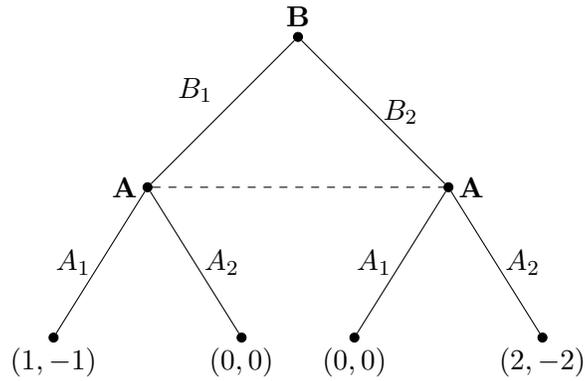
\begin{figure}[!t]	%[h]
    \centering
\begin{tikzpicture}[grow=down]
      \node(0)[player]{} 
      child{node(1)[player] {} 
        child{node[player] {} edge from parent node [left]{$A_1$} 
        }
        child{node[player] {} edge from
          parent node [right] {$A_2$} 
        } 
        edge from parent node [above left] {$B_1$} 
      } 
      child{node(2)[player] {}
        child{node[player] {} edge from parent node [left] {$A_1$} 
        }
        child{node[end] {} edge from parent node [right] {$A_2$} 
        } 
        edge from parent node [right] {$B_2$} 
      };

      %% information set
 	\draw[dashed](1)to(2);

      %% specify players
      \node[above]at(0){$\mathbf{B}$}; 
      \node[right]at(2){$\mathbf{A}$};   
      \node[left] at (1) {$\mathbf{A}$}; 

      %% specify payoffs
 \payoff{1-1}1{-1}
 \payoff{1-2}00
 \payoff{2-1}00
 \payoff{2-2}2{-2} 

    \end{tikzpicture}
   
\caption{Matching Pennies Extensive form game}
\label{f-example}
  \end{figure}
%%===================================

% AB - using the rule "oracle" as the noun, and "Oracle Games" as what we study (like Prisoner's Dilemma)

Since every simultaneous game is equivalent to a sequential game in which neither player observes the actions taken by the other (see e.g.~\citep{GonzalesDiaz2010}), we first consider this as a sequential game in which player B selects a strategy first, as illustrated in Fig \ref{f-example}. The mixed strategy Nash equilibrium of this game gives a probability over these pure strategies shown in the tree, while in any actual realization of this game, the players will randomly play one of these strategies determined by a random process.

We next introduce an oracle, a third player who has access to the realized pure strategy which will actually be played by player B.  
We modify the game by inserting three additional stages into the standard sequence as follows:
\begin{enumerate}
\item  Player A chooses a nonnegative amount $x$ to pay to the oracle.
\item  Player B chooses a strategy.
\item  %The oracle performs a random operation and w
With probability $I(x)$ the oracle informs player A of player B's realized pure strategy (respond), and with probability $1-I(x)$ remains silent (silent).
\item Player A then chooses a strategy and the game resolves, with player A's final payoff being decreased by the payment $x$ chosen earlier.
\end{enumerate}
Fig \ref{BFirst} shows the extensive form of this game. Note that since no information is given to player B at any point, the order of stages 1-3 may be rearranged in several ways, which allows for easier analysis without affecting the game.

%oracle diagram VERSION WITH B FIRST
%%===================================
% tree layout
\tikzstyle{level 1}=[level distance=20mm, sibling distance = 5mm]
\tikzstyle{level 2}=[level distance=20mm, sibling distance = 40mm]
\tikzstyle{level 3}=[level distance=20mm, sibling distance = 20mm]
\tikzstyle{level 4}=[level distance=20mm, sibling distance = 10mm]
% node styles
\tikzstyle{player} = [circle, draw, inner sep=1.2, fill=black]
\tikzstyle{end} = [circle, draw, inner sep=1.2, fill=black ]
\tikzstyle{hollow} =[circle,draw,inner sep=0.0,fill=white]

\newcommand{\singlePayoff}[3][below]{\node[#1] at (#2) {$#3$};}

 \begin{figure}[!t]	%[h]
    \centering
\begin{tikzpicture}[grow=down]
  \node(0)[player]{}
     child{node(a)[hollow, white]{} edge from parent node [left] {$$} }
     child{[white] node(1)[player, yshift = -5]{} 
      child{[black] node(2)[player] {} 
      	child{node[player] {}
		child{node[player] {} edge from parent node [left]{$A_1$}   }
		child{node[player] {} edge from parent node [right]{$A_2$}   }
	 edge from parent node [left]{$respond$}   }
      	child{node[player] {}
		child{node[player] {} edge from parent node [left]{$A_1$}   }
		child{node[player] {} edge from parent node [right]{$A_2$}   }
 	edge from parent node [right] {$silent$}     } 
      edge from parent node [left] {$B_1$}    } 
      child{[black] node(3)[player] {}
      	child{node[player] {}
		child{node[player] {} edge from parent node [left]{$A_1$}   }
		child{node[player] {} edge from parent node [right]{$A_2$}   }
	 edge from parent node [left] {$respond$}     }
      	child{node[end] {}
		child{node[player] {} edge from parent node [left]{$A_1$}   }
		child{node[player] {} edge from parent node [right]{$A_2$}   }
	 edge from parent node [right] {$silent$}      } 
      edge from parent node [right] {$B_2$}   }
  edge from parent node [right] {$$} }
 child{node(b)[hollow, white]{} edge from parent node [right] {$x$} }

;

      %% information set
 	
	\draw[bend right, dashed](2-2)to(3-2);
	\draw[bend right](a)to(b);

      %% specify players
      \node[above]at(0){$\mathbf{A}$}; 
      \node[right, xshift = 20]at(1){$\mathbf{B}$};   
      \node[right] at (3) {$\mathbf{Oracle}$}; 
      \node[right] at (3-2) {$\mathbf{A}$}; 

      %% specify payoffs
\singlePayoff{2-1-1}1
\singlePayoff{2-1-2}0
\singlePayoff{2-2-1}1
\singlePayoff{2-2-2}0
\singlePayoff{3-1-1}0
\singlePayoff{3-1-2}2
\singlePayoff{3-2-1}0
\singlePayoff{3-2-2}2

    \end{tikzpicture}
   
    \caption{Game tree illustrating the initial inclusion of an oracle providing information to player A.}
 \label{BFirst}
  \end{figure}
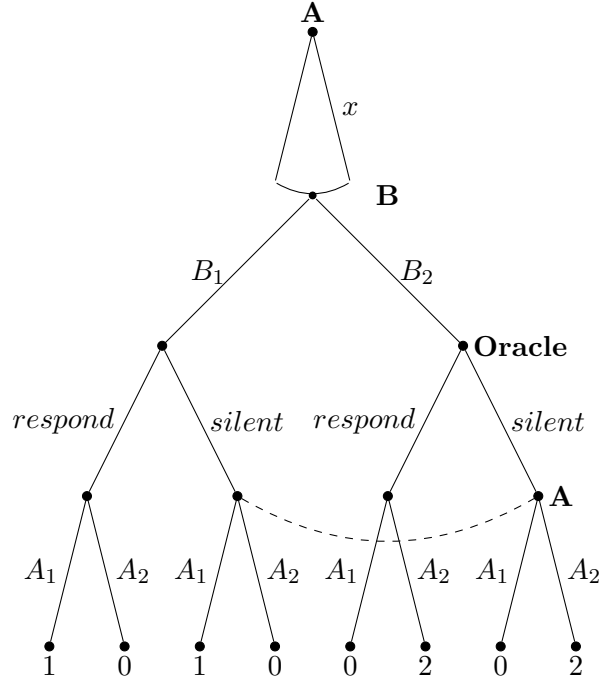
%%===================================

First note that in this example, player A's best response to each of player B's strategies is unique (we will restrict ourselves to games with this property for the remainder of the paper).  Therefore, in the case that the oracle responds and provides B's strategy, the rational response of player A is already determined.  Thus player A only makes two choices: the amount $x$ of payment to the oracle, and the strategy choice when the oracle does not respond.  
Thus we may equivalently consider a sequence in which A makes a tentative decision of what to play at the beginning of the game, and changes her mind only if the oracle responds:  
%The following structure leads to equivalent behavior and payoffs for every payoff matrix:

\begin{enumerate}
\item  Player A chooses any nonnegative amount $x$ to pay to the oracle.
\item  %The oracle performs a random operation and w
With probability $I(x)$ the oracle commits to informing player A of B's strategy at a later time, and with probability $1 - I(x)$ commits to remaining silent.
\item  Player A tentatively chooses a strategy to play if not given a response.
\item Player B chooses a strategy to play.
\item  If the oracle committed to respond, it does so now, Player A ignores his previous choice and chooses the best response to player B's strategy.  If the oracle committed to remaining silent, then player A uses the tentative choice.  In either case, the game resolves and player A's payoff is reduced by $x$.

\end{enumerate}

%oracle diagram VERSION WITH OAB
%%===================================
% tree layout
\tikzstyle{level 1}=[level distance=20mm, sibling distance = 5mm]
\tikzstyle{level 2}=[level distance=20mm, sibling distance = 40mm]
\tikzstyle{level 3}=[level distance=20mm, sibling distance = 20mm]
\tikzstyle{level 4}=[level distance=20mm, sibling distance = 10mm]
% node styles
\tikzstyle{player} = [circle, draw, inner sep=1.2, fill=black]
\tikzstyle{end} = [circle, draw, inner sep=1.2, fill=black ]
\tikzstyle{hollow} =[circle,draw,inner sep=0.0,fill=white]

 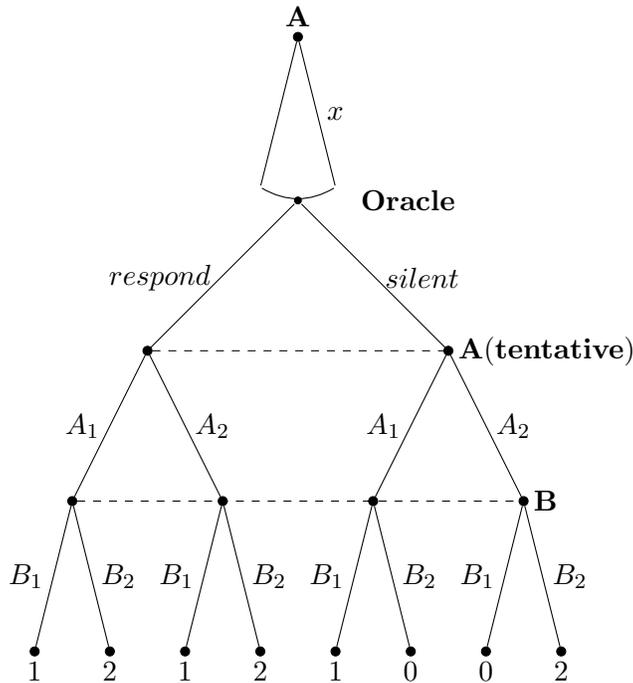
\begin{figure}[h]
    \centering
\begin{tikzpicture}[grow=down]
  \node(0)[player]{}
     child{node(a)[hollow, white]{} edge from parent node [left] {$$} }
     child{[white] node(1)[player, yshift = -5]{} 
      child{[black] node(2)[player] {} 
      	child{node[player] {}
		child{node[player] {} edge from parent node [left]{$B_1$}   }
		child{node[player] {} edge from parent node [right]{$B_2$}   }
	 edge from parent node [left]{$A_1$}   }
      	child{node[player] {}
		child{node[player] {} edge from parent node [left]{$B_1$}   }
		child{node[player] {} edge from parent node [right]{$B_2$}   }
 	edge from parent node [right] {$A_2$}     } 
      edge from parent node [left] {$respond$}    } 
      child{[black] node(3)[player] {}
      	child{node[player] {}
		child{node[player] {} edge from parent node [left]{$B_1$}   }
		child{node[player] {} edge from parent node [right]{$B_2$}   }
	 edge from parent node [left] {$A_1$}     }
      	child{node[end] {}
		child{node[player] {} edge from parent node [left]{$B_1$}   }
		child{node[player] {} edge from parent node [right]{$B_2$}   }
	 edge from parent node [right] {$A_2$}      } 
      edge from parent node [right] {$silent$}   }
  edge from parent node [right] {$$} }
 child{node(b)[hollow, white]{} edge from parent node [right] {$x$} }

;

      %% information set
 	\draw[dashed](2)to(3);
	\draw[dashed](2-1)to(3-2);
	\draw[bend right](a)to(b);

      %% specify players
      \node[above]at(0){$\mathbf{A}$}; 
      \node[right, xshift = 20]at(1){$\mathbf{Oracle}$};   
      \node[right] at (3) {$\mathbf{A (tentative)}$}; 
      \node[right] at (3-2) {$\mathbf{B}$}; 

      %% specify payoffs
\singlePayoff{2-1-1}1
\singlePayoff{2-1-2}2
\singlePayoff{2-2-1}1
\singlePayoff{2-2-2}2
\singlePayoff{3-1-1}1
\singlePayoff{3-1-2}0
\singlePayoff{3-2-1}0
\singlePayoff{3-2-2}2

    \end{tikzpicture}
   
    \caption{Standard game tree construction for an Oracle Game.}
 \label{OAB}
  \end{figure}
%%===================================

Fig \ref{OAB} shows the extensive game for this version. We model the subgame at stage 2 as a Bayesian game with two possible states \citep{GonzalesDiaz2010}.  When the oracle does not respond, the payoff matrix for the Oracle Game is the same as the game without an oracle.  When the oracle does respond, the payoffs for each player are given by player A's best response in the column determined by player B's choice (since $x$ is constant in this subgame, we omit it in the payoff matrices as it does not affect equilibria).  This is represented as the matrix $R$:
\begin{table}[H]
\renewcommand{\arraystretch}{1.1}
\Large
\centering
  \begin{tabular}[h]{r|c|c|c|}
 \multicolumn{1}{c}{\diagbox[height=1.9em]{\normalsize 1}{\normalsize
    ~2}} & \multicolumn{1}{c}{$B_1$} & \multicolumn{1}{c}{$B_2$}  \\\cline{2-3}
    $A_1$   & $1,-1$ & $2,-2$ \\\cline{2-3}
    $A_2$ &  $1,-1$ &  $2,-2$ \\ \cline{2-3}
  \end{tabular}
\end{table}
\noindent
where $A_1$ and $A_2$ are the tentative decisions for player A.  We refer to $R$ as the maximal matrix of the game, since the payoffs for player A are equal to the maximum in each column of the original payoff matrix.  Since this matrix shows the payoffs when the oracle does respond, it is natural that the payoffs in each column are identical since he changes his mind and ignores his previous decision.  If $M$ is the original payoff matrix, then the matrix of the expected values that the players perceive in the subgame is given by $M \cdot (1 - I(x)) + R \cdot I(x)$.  In this example, it becomes:

\begin{table}[H]
\renewcommand{\arraystretch}{1.1}
%\Large
\centering
  \begin{tabular}[h]{r|c|c|c|}
 \multicolumn{1}{c}{\diagbox[height=1.9em]{\normalsize 1}{\normalsize
    ~2}} & \multicolumn{1}{c}{$B_1$} & \multicolumn{1}{c}{$B_2$}  \\\cline{2-3}
    $A_1$   & $1,-1$ & $2I(x),-2I(x)$ \\\cline{2-3}
    $A_2$ &  $I(x), -I(x)$ &  $2,-2$ \\ \cline{2-3}
  \end{tabular}
\end{table}

The equilibria of this game will depend on the value of the oracle function $I(x)$, which will depend on the value $x$ paid by Player A.  The strategy space may be described as $S = \{s_a, s_b, x\}$ where $s_b$ is B's strategy, $s_a$ is A's tentative strategy, and $x$ is A's payment to the oracle.  To simplify notation, we will often use $I$ to denote $I(x)$, and likewise $I'$ denotes $\frac{dI}{dx}$, understood to be evaluated at the value of $x$ being played by A.

One of our main results for these oracle games are that transitions occur at critical values of the purchased probability $I(x)$. 
For any $a, b \in \mathbb{R}$ we define $x_a$ to be the smallest payment $x$ such that $I(x) = a$, and $y_b$ to be any $x$ such that $I'(x) = b$.  
\endnote{Since $I(x)$ is weakly concave down, $I'(x)$ may be constant and equal to $b$ on some interval.  If so, then $y_b$ can refer to any one of the values on that interval, and any statement we make about $y_b$ is true for all such values.  When this occurs in a Nash Equilibrium, the Oracle Game will have multiple equilibria, one for each choice of $y_c$}
For  the game considered in this example, we classify the equilibria into one of three cases, depending only on properties of the oracle function $I(x)$:

\medskip

\noindent
{\it Case 1:}
If $I'(0) \leq \frac{3}{2}$, the equilibrium is $\{(\frac{2}{3},\frac{1}{3}),(\frac{2}{3},\frac{1}{3}),0\}$, since player A pays $x = 0$, the players behave as they would if there was no oracle. 

\medskip

\noindent
{\it Case 2:}
If $ I'(0) \geq \frac{3}{2} \geq I'(x_\frac{1}{2})$, the equilibrium is $\{(\frac{2 - I}{3(1 - I)},\frac{1-2I}{3(1-I)}),(\frac{2}{3},\frac{1}{3}),y_\frac{3}{2}\}$. 

\medskip

\noindent
{\it Case 3:} If $I'(x_\frac{1}{2}) \geq \frac{3}{2}$, the equilibrium is $\{(1,0),(\frac{2I' - 1}{2I'}, \frac{1}{2I'}),x_\frac{1}{2}\}$. 

\medskip

\begin{figure}[!b]
\centering
\includegraphics[scale = 0.35]{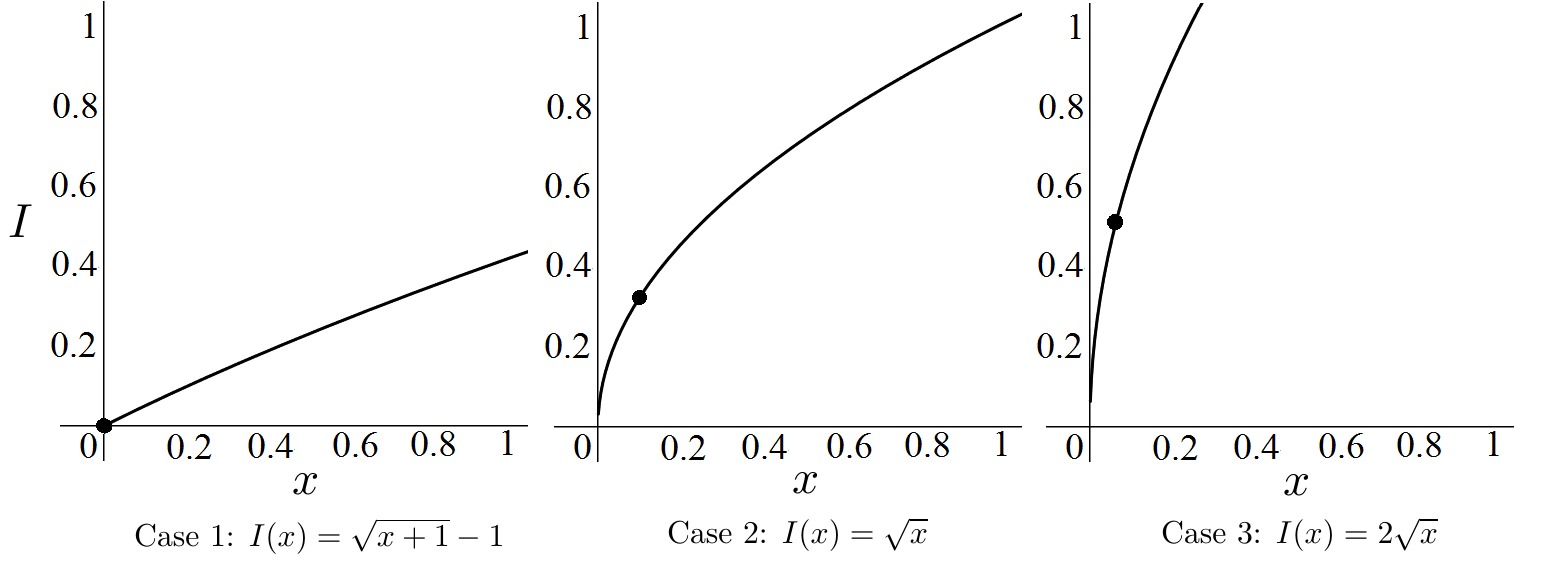}
\put(-380,120){(a)}
\put(-250,120){(b)}
\put(-121,120){(c)}
%\put(-420,80){$I(x)$}
%\put(-20,4){$x$}
\caption{Equilibrium payments (black dot) for the game in Example 1, shown for oracle functions
(a) $I(x) = \sqrt{x+1}-1$;
(b) $I(x) = \sqrt{x}$;
(c) $I(x) = 2\sqrt{x}$.
}
\label{examplegraphs}
\end{figure}

\noindent
The determination of these equilibria follows from Theorem \ref{Equilibrium Conditions}, presented below in Section \ref{mainresults}.  Figure \ref{examplegraphs} illustrates oracle functions leading to each of these cases for this example.  Note that all $I(x)$ are appropriately capped at $I = 1$, as it is a probability.

%%===================================
%%===================================

\newpage
\subsection{Motivating Example 2}

We next consider a $3 \times 3$ symmetric game, defined by the matrix:

\begin{table}[H]
\renewcommand{\arraystretch}{1.1}
\Large
\centering
  \begin{tabular}[h]{r|c|c|c|}
 \multicolumn{1}{c}{\diagbox[height=1.9em]{\normalsize 1}{\normalsize
    ~2}} & \multicolumn{1}{c}{$B_1$} & \multicolumn{1}{c}{$B_2$} 
        &\multicolumn{1}{c}{$B_3$} \\\cline{2-4}
    $A_1$   & $1,-1$ & $0,0$ & $0,0$  \\\cline{2-4}
    $A_2$ &  $0,0$ & $2,-2$ & $0,0$ \\ \cline{2-4}
    $A_3$ &  $0,0$ & $0,0$ & $4,-4$ \\ \cline{2-4}
  \end{tabular}
\end{table}

\noindent
Note that this matrix contains the previous example as a submatrix.  With no oracle, the only equilibrium is when A and B both play the mixed strategy $(\frac{4}{7}, \frac{2}{7}, \frac{1}{7})$.  If Player A is given access to an oracle, then using the same process as before, the matrix becomes:

\begin{table}[H]
\renewcommand{\arraystretch}{1.1}
\Large
\centering
  \begin{tabular}[h]{r|c|c|c|}
 \multicolumn{1}{c}{\diagbox[height=1.9em]{\normalsize 1}{\normalsize
    ~2}} & \multicolumn{1}{c}{$B_1$} & \multicolumn{1}{c}{$B_2$} 
        &\multicolumn{1}{c}{$B_3$} \\\cline{2-4}
    $A_1$   & $1,-1$ & $2I,-2I$ & $4I,-4I$  \\\cline{2-4}
    $A_2$ &  $I,-I$ & $2,-2$ & $4I,-4I$ \\ \cline{2-4}
    $A_3$ &  $I,-I$ & $2I,-2I$ & $4,-4$ \\ \cline{2-4}
  \end{tabular}
\end{table}

\noindent
For this game, the equilibria fall into one of the following cases:

\medskip

Case 1: If $I'(0) \leq \frac{7}{8}$ the equilibrium is $\{(\frac{4}{7}, \frac{2}{7}, \frac{1}{7}),(\frac{4}{7}, \frac{2}{7}, \frac{1}{7}),0\}$. \\

Case 2: If $I'(0) \geq \frac{7}{8} \geq I'(x_\frac{1}{5})$, the equilibrium is 
$$
\left \{ \left(\frac{4+I}{7(1-I)},\frac{2-3I}{7(1-I)},\frac{1-5I}{7(1-I)}\right), \left(\frac{4}{7}, \frac{2}{7}, \frac{1}{7}\right), y_\frac{7}{8} \right\}
$$

At I = $\frac{1}{5}$, the probability of $A_3$ reaches 0, where A can no longer maintain B's indifference since $B_3$ is now weakly dominated by a mixed strategy of $B_1$ and $B_2$.

Case 3: If $\frac{7}{8} \leq I'(x_\frac{1}{5}) \leq \frac{3}{2}$, the equilibrium is 
$$
\left \{ \left (\frac{3}{4},\frac{1}{4},0 \right), \left(\frac{8I'-2}{10I'},\frac{4I'-1}{10I'},\frac{3-2I'}{10I'}\right), x_\frac{1}{5} \right \}
$$

%%===================================
\begin{figure}[!t]
\centering
\includegraphics[width=13cm]{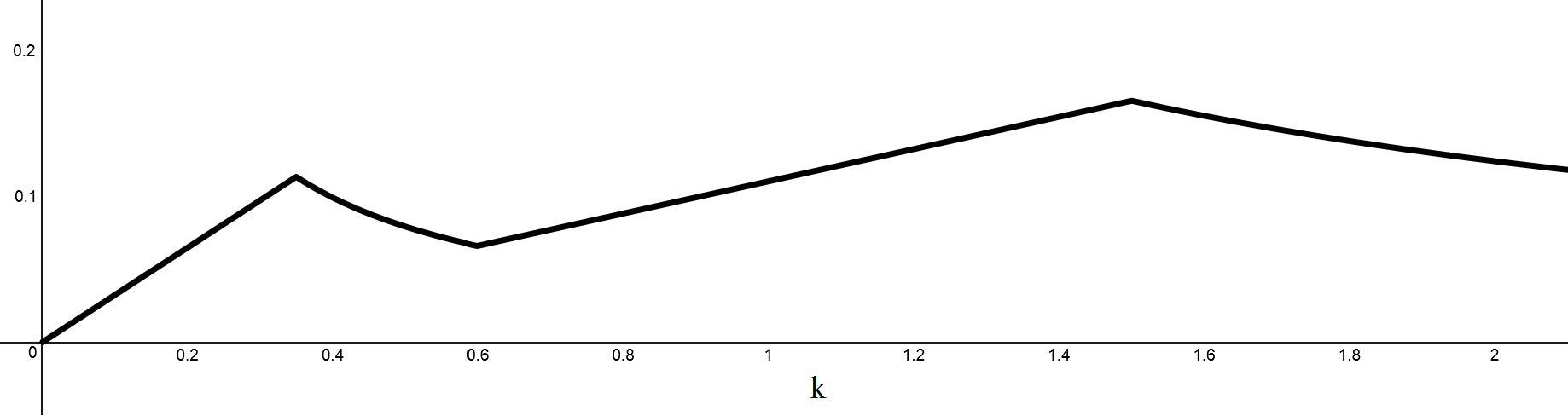}
	\put(-380,75){$x_e$}
	\put(-340,85){2}
	\put(-280,85){3}
	\put(-180,85){4}
	\put(-50,85){5}
	\multiput(-300,0)(0,8){13}{\line(0,3){2}}
	\multiput(-257,0)(0,8){13}{\line(0,3){2}}
	\multiput(-103,0)(0,8){13}{\line(0,3){2}}
	%\put(-60,-10){(b)}

\vspace{3mm}

\includegraphics[width=13cm]{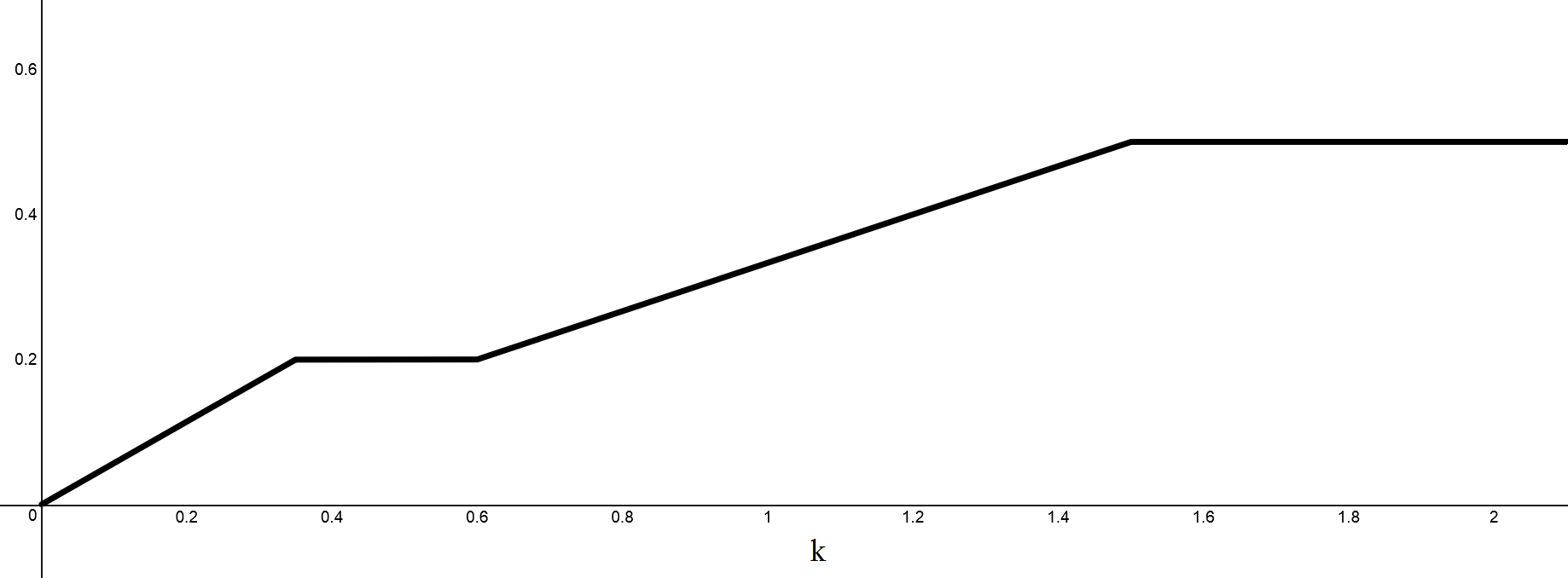}
	\put(-391,100){$I(x_e)$}
	\multiput(-300,0)(0,8){18}{\line(0,3){2}}
	\multiput(-257,0)(0,8){18}{\line(0,3){2}}
	\multiput(-103,0)(0,8){18}{\line(0,3){2}}

\caption{Equilibrium amount $x_e$ paid to the oracle by player A (top) for the game in Example 2, 
and the resulting response rate $I(x_e)$ (bottom)  
as functions of the parameter $k$ for the oracle response probability function $I(x) = \sqrt{kx}$. %RPF - reponse probability function
}
\label{IKXK}
\end{figure}
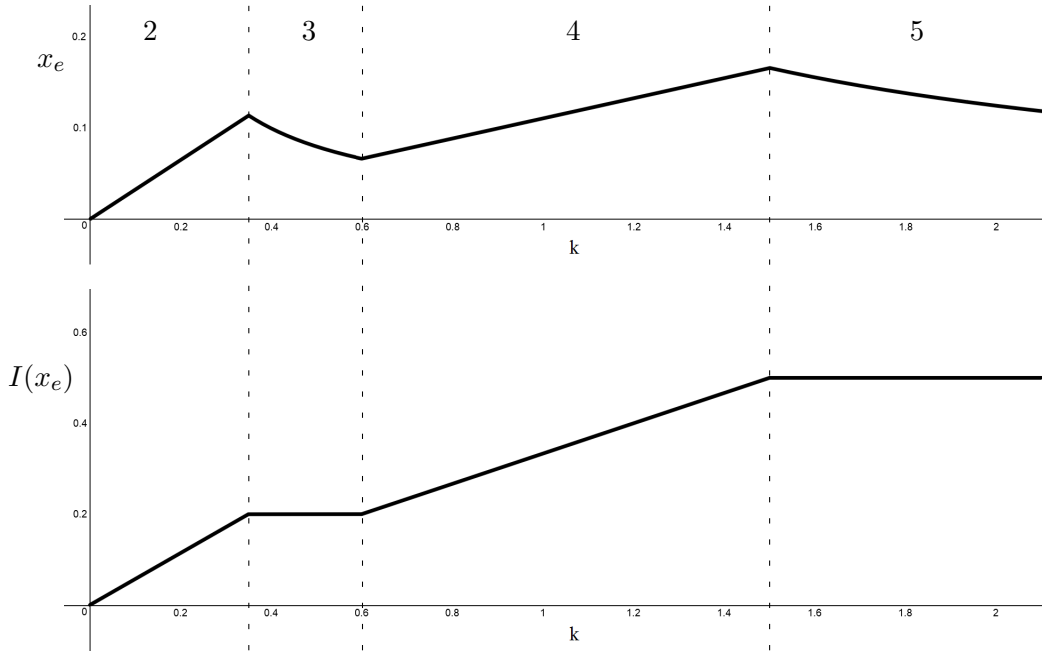	
%%===================================

At $I' = \frac{3}{2}$, the probability of  $B_3$ reaches 0, and B can no longer prevent A from increasing $x$.

At this point, A and B have both eliminated strategy 3 (B will never play it as it is now a dominated strategy, and A will never play it since after eliminating column 3 it is also dominated).  Thus the game reduces to the matrix:

\begin{table}[H]
\renewcommand{\arraystretch}{1.1}
\Large
\centering
  \begin{tabular}[h]{r|c|c|c|}
 \multicolumn{1}{c}{\diagbox[height=1.9em]{\normalsize 1}{\normalsize
    ~2}} & \multicolumn{1}{c}{$B_1$} & \multicolumn{1}{c}{$B_2$}  \\\cline{2-3}
    $A_1$   & $1,-1$ & $2I,-2I$ \\\cline{2-3}
    $A_2$ &  $I,-I$ &  $2,-2$ \\ \cline{2-3}
  \end{tabular}
\end{table}

Note that this is identical to the matrix from the first game, thus, all equilibria that come from this matrix will be identical.

Case 4: If $I'(x_\frac{1}{5}) \geq \frac{3}{2} \geq  I'(x_\frac{1}{2})$, the equilibrium is

$$
 \left\{ \left(\frac{2 - I}{3(1 - I)}, \frac{1 - 2I}{3(1 - I)}, 0 \right), \left(\frac{2}{3},\frac{1}{3}, 0 \right), y_\frac{3}{2} \right \}
$$

Case 5: If $I'(x_\frac{1}{2}) \geq \frac{3}{2}$, the equilibrium is
$$
 \left\{ \left(1,0,0 \right), \left(\frac{2I' - 1}{2I'}, \frac{1}{2I'}, 0 \right), x_\frac{1}{2} \right \}
$$

In general, if $G$ is a game, and $H$ is a game which contains $G$ as a subgame, then if $I(x)$ is an oracle function that causes all strategies in $H$ which are not in $G$ to become dominated, then the equilibria of games $G$ and $H$ subject to oracle $I$ will be the same.

To illustrate how the amount paid by Player A at equilibrium $x_e$ varies as the cost of information decreases in this example, we consider a one parameter family of oracle functions $I(x) = \sqrt{kx}$ with parameter $k$; the cost of information decreases as $k$ increases. 
Figure \ref{IKXK} shows the dependent of The dependence of $x_e$ on $k$, as well as the changes in the purchased probability of response $I(x_e)$. 

The numbers between the dotted lines indicate which case the equilibrium corresponds to in that interval; note that case 1 does not occur for any $k$ since $\sqrt{kx}$ has infinite slope at $x = 0$.  For Cases 2 and 4, Player A gradually increases $x$ in response to the cheaper information, while maintaining B's indifference by adjusting $s_a$ to compensate.  In cases 3 and 5, A maintains $I$ at a constant value (which costs less to maintain as information becomes cheaper), and B maintains A's indifference by adjusting $s_b$ away from exploitable strategies.

In the following sections we prove results about Oracle Games indicating that most well-behaved games have equilibria similar to these example cases (for certain notions of ``well-behaved" and ``similar"), and we show how these equilibria are determined.

%%%%%%%%%%%%%%%%%%%%%%%%%%%%%%%%
\subsection{General Definitions for Oracle Games}

Let $G$ be a simultaneous, two-player game with the $m \times n$ payoff matrix $M$ and players A and B.  Let $\gi$ be the game where A and B play game $G$ but A is given access to an oracle with function $I(x)$.  If A's maximal payoff in each column of $M$ is unique, then A's best response to an oracle response is predetermined, which means that A does not have to specify a strategy choice when the oracle responds. 
The set of strategy profiles is then expressed as 
$S = \{s_a, s_b, x\}$ where $s_b$ is B's strategy, $s_a$ is A's strategy when the oracle does not respond, and $x$ is A's payment to the oracle.  We make no meaningful distinction between pure and mixed strategies, except to note that an assumption we make later allows all oracle payments $x$ to be considered as pure strategies.

For each $j$, let $\alpha _j$ be the index of the row corresponding to the highest payoff to A in column $j$ of $M$ (we assume this is unique for each $j$).   We define the maximal matrix $R$ by $R_{i,j} =M_{\alpha _j, j}$, such that every outcome in $R$ is a copy of the outcome in $M$ corresponding to A's best response to strategy $j$.  Let $C$ be the $m \times n$ matrix where the payoff to A is $1$ and the payoff to B is $0$ in every cell.  Then for every $x \in [0, \infty)$, A paying the oracle $x$ induces a Bayesian game with expected payoffs  
$$
M \cdot (1 - I(x)) + R \cdot I(x) - C \cdot x.
$$  
Let $\mix{x} := M \cdot (1 - I(x)) + R \cdot I(x)$.  Since the equilibria of a payoff matrix do not change with a constant reduction in all of the payoffs for either player, for each fixed $x$ this will have the same equilibria as the actual induced payoff matrix.

Then $s \in S$ is a Nash equilibrium if and only if A and B are both indifferent on changing each of their strategies.  Thus a necessary condition for an equilibrium must be that for whichever $x$ player A is paying, ${s_a, s_b}$ must be an equilibrium for the matrix $\mix{x}$, since otherwise A or B could profit by changing their strategies.
We also write the expected payoff $E_a$ of A playing $\gi$ in terms of 
A's expected payoff $E_r$ from $\gi$ given a response and the expected payoff $E_n$ given no response, as 
$$
E_a(s_a, s_b, x) = E_n(s_a,s_b) \cdot (1 - I) + E_r(s_b) \cdot I - x.
$$

\begin{Definition}
We define the Value of Information $V$ to be the marginal increase in Player A's expected payoff with increasing probability of response $I$, that is 
$$
V :=\frac{ \partial E_a }{ \partial I} =  E_r - E_n.
$$  
\end{Definition}

\noindent
Thus the value of information is the change in expected benefit for A due to receiving ``more response", i.e.~a greater chance of correct response.  If we assume that player A always chooses the optimal $s_a$ for the particular cross section $\mix{x}$, then $V$ can be expressed solely as a function of $s_b$.   It immediately follows that $V \geq 0$ for all $s_a, s_b$, since A's payoff when the oracle responds is always at least as good as her payoff when it is silent.  

\noindent
{\bf Remarks:}

1) whenever $s_b$ is a pure strategy, A will play the best response to $s_b$, regardless of whether the oracle responds or not. Thus $E_r = E_n$, and $V = 0$, which makes sense in this case where the information has no value.

\smallskip

2)  $E_a$ is linear with respect to $s_b$: if $s_1$ and $s_2$ are strategies for B, and $p \in [0,1]$, then $E_a(s_a, ps_1 + (1-p)s_2, x) = pE_a(s_a,s_1, x) + (1-p)E_a(s_a,s_2, x)$.  This also implies $V$ is linear with respect to $s_b$.

\begin{Definition}
We say that $\gi$ has a node at $c$ if one of B's strategies changes from dominated to undominated (or vice versa) in $\mix{x}$ at $x = c$.
\end{Definition}

\noindent
Such nodes are represented at equilibrium for Example 2 by the dashed lines in Fig.~\ref{IKXK}.

%We observe that case 3 in the first example and cases 3 and 5 in the second example above correspond to equilibria with oracle payment $x$ at a node.

\section{Fundamental Properties of Oracle Games}

We first derive some fundamental results that elucidate the basic properties of these Oracle Games.

 \begin{Proposition}
 \label{PureStrategyEquivalence}
 If $\{s_a, s_b\}$ is a pure strategy Nash equilibrium of G, then $\{s_a, s_b, 0\}$ is a Nash equilibrium of $\gi$.
 \end{Proposition}

 \textbf{Proof:} 
If $\{s_a,s_b\}$ is a pure strategy Nash equilibrium in $G$, then $s_a$ is a best response to $s_b$, and $s_b$ is a best response to $s_a$ in $M$, and if $x = 0$ then the oracle never responds, so $\mix{0} = M$.  And since B is playing a pure strategy, $s_a$ will be a best response to $s_b$ regardless of whether the oracle responds or not, so A cannot benefit by increasing $x$.  Thus, no player has an incentive to change their strategies in any way, and $\{s_a, s_b, 0\}$ is a Nash equilibrium of $\gi$.
 $\blacksquare $

In other words, the oracle does not affect pure strategy equilibria.  This is natural, since in a pure strategy equilibria, both players are playing pure strategies, so information confirming what is already known adds no value. \\

%%%%%%%%%%%%%%%%
\begin{Definition}
We define two oracle functions $I(x)$ and $J(x)$ to be equivalent ($I \cong J$) if for every game $G$, the set of equilibrium strategies (excluding the oracle payment) and resulting expected payoffs (including the payment) are identical for $\gi$ and $G \vert J$.
\end{Definition}
%%%%%%%%%%%%%%%%

\noindent
This definition is useful because of the following results, which are based on the fact that a rational player will never pay {\it more} for {\it less} information (or in our case, for a less probable response). 

 \begin{Proposition}
 \label{OracleEquivalence}
 Every oracle function is equivalent to one which is continuous, nondecreasing, and (weakly) concave down.
 \end{Proposition}

\textbf{Proof:} 
Given any oracle function $I(x)$, we will construct another oracle function $J(x)$ based on $I$ such that $J$ is continuous, nondecreasing and (weakly) concave down, and then show that $J$ is equivalent to $I$.  

We first construct a nondecreasing version of $I$. 
Suppose that there exists some $c_2 >c_1$ such that $I(c_2) < I(c_1)$; player A will never pay $c_2$ since it's dominated by $c_1$ (choosing $c_2$ over $c_1$ means paying more for less information).
The value of information is always nonnegative, therefore A's expected value must be nondecreasing with respect to $I$, and strictly decreasing with respect to $x$.  If we let $J_1$ be an oracle function with 
$$
J_1(x) = \sup(I(a) : a \leq x)
$$ 
then $J_1$ is nondecreasing since it's taking the supremum over a growing set. And $J_1$ is equivalent to $I$ because any values of $x$ that differ between $I$ and $J_1$ are ones for which $I$ has dropped below $\sup(I)$, which are also $x$ that  A would never pay.  Similarly in $G \vert J_1$, A will also never pay them because that would be paying more for the same amount of information.

The fact that Player A can play a mixed strategy between two oracle payments leads to the second result, which we show by constructing a non-concave up version of $J_1$. 
Let $c_1$ and $c_2$ be any numbers in $[0,\infty)$, and A's mixed strategy be to pay $c_1$ with probability $p$ and $c_2$ with probability $(1 - p)$.  The expected amount A will pay is then 
$pc_1 + (1 - p)c_2 = \bar{x}$, 
and the expected probability that the oracle will respond will be $pI(c_1) + (1 - p)I(c_2) = \bar{I}$. The combination of these two yields the same results as another oracle function which took on the value $\bar{I}$ at the point $\bar{x}$. 
Thus the oracle function $J_1$ is equivalent to the supremum of its convex hull: 
$$
J(x) = \sup( pJ_1(c_1) + (1 - p)J_1(c_2) )
$$ 
where the supremum is over all $c_1$ and $c_2$ in $[0,\infty)$ and all $p$ in [0, 1].  The supremum of the convex hull of any function is automatically continuous and (weakly) concave down. 
Note also that  $J$ is nondecreasing because $J_1$ is.
$\blacksquare $

	%%===================================
	\begin{figure}[!t]
	\center{
	\includegraphics[width=5cm]{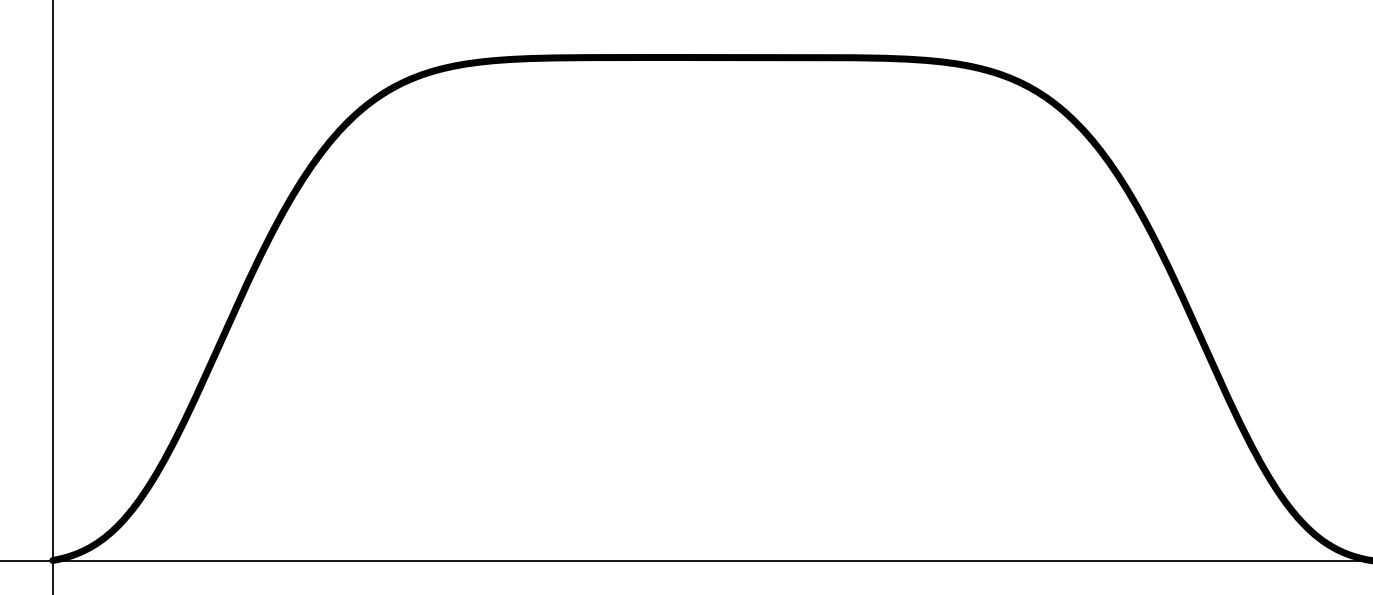}
	%\put(-260,80){(a)}
	\Large \put(-160,45){$I$}
	}
	\vspace{3mm}
	\center{
	\includegraphics[width=5cm]{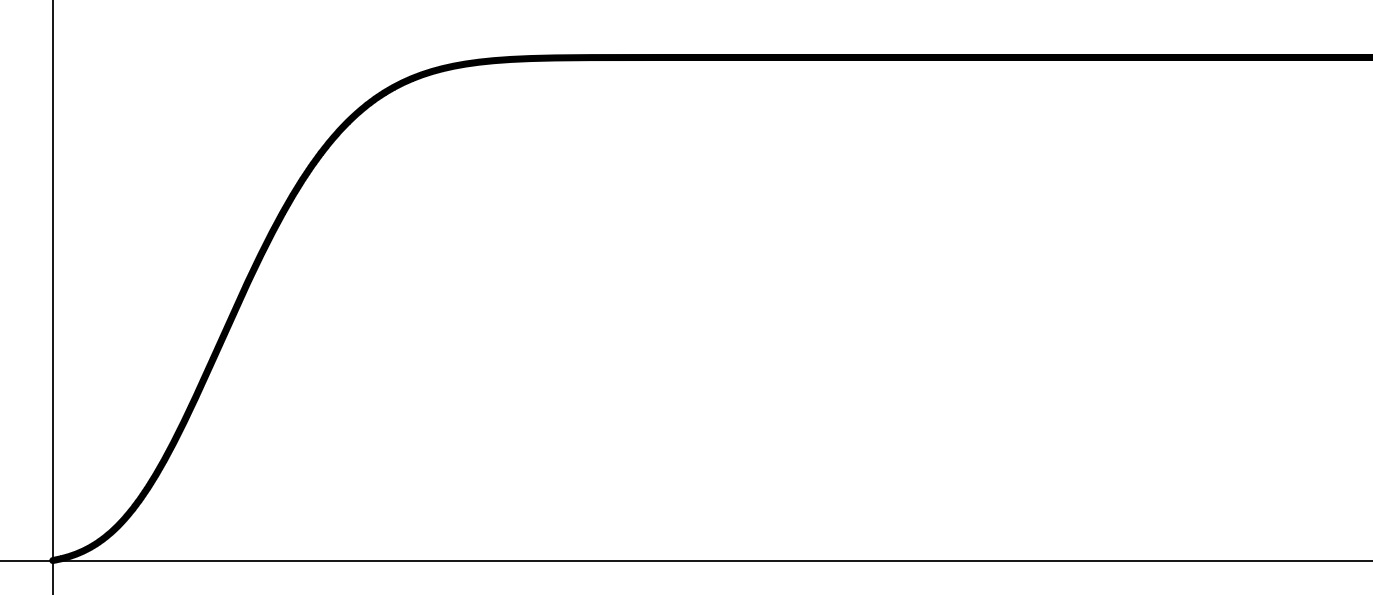}
	%\put(-260,80){(b)}
	\Large \put(-160,45){$J_1$}
	}
	\vspace{3mm}
	\center{
	\includegraphics[width=5cm]{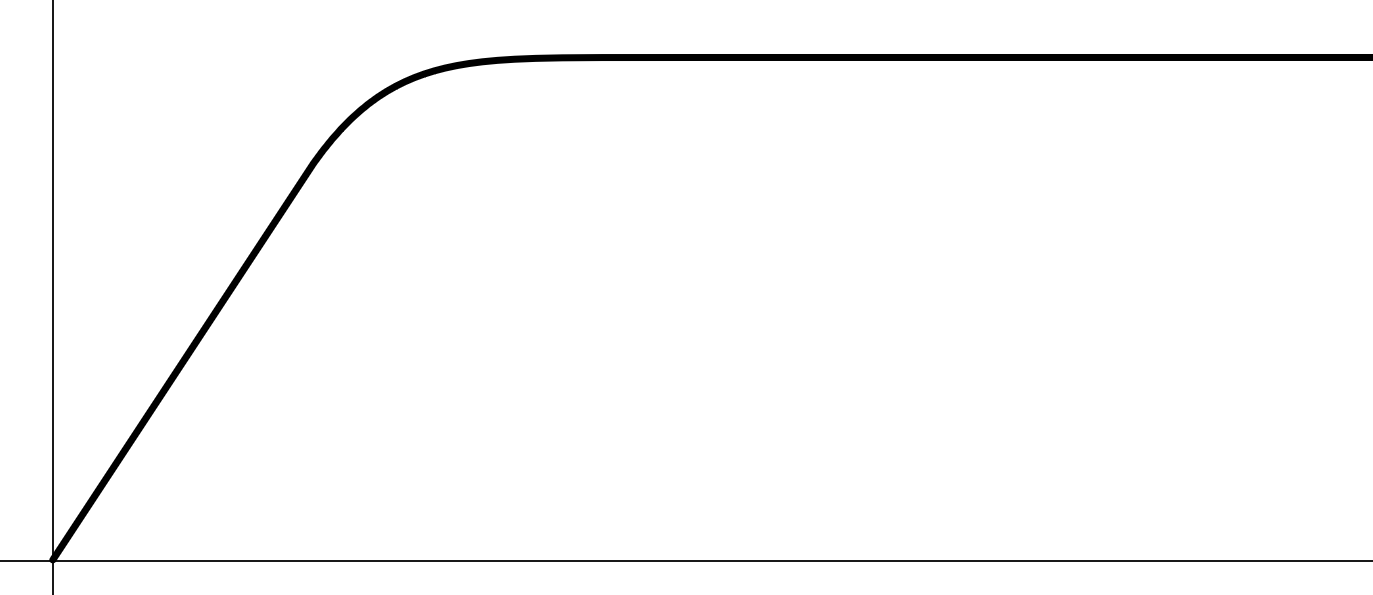}
	%\put(-260,80){(c)}
	\Large \put(-160,45){$J$}
	\large \put(2,-5){$x$}
	}
	\caption{Illustration of the equivalence proven in Proposition \ref{OracleEquivalence} (see text): 
	the original oracle function $I(x)$ (top); 
	a nondecreasing but equivalent oracle function $J_1(x)$ (middle); 
	the final nondecreasing and weakly concave down oracle function $J(x)$ (bottom). 
	}
	\label{OracleMod}
	\end{figure}	
	%%===================================

In Figure \ref{OracleMod} we show an example of this construction process for a particular 
$I(x)$ (Fig.~\ref{OracleMod}a), with equivalent nondecreasing oracle functions $J_1(x)$ (Fig.~\ref{OracleMod}b), and the full simplification of the proposition, Fig.~\ref{OracleMod}c. We next show that any $\gi$ with a nonzero cost of zero information can be shifted to an equivalent game for which $I(0) = 0$.

\begin{Proposition}
\label{I starts at 0}
Suppose $\gi$ is a game with payoff matrix $M$ and oracle function $I(x)$ with $I(0) > 0$.  Then there exist a game H and oracle function J(x) with J(0) = 0 such that $\gi \cong H \vert J$.
\end{Proposition}

For $\gi$, let $I(0) = c > 0$, and note that $c \le 1$.  Define the following
$$
N = M_{I(0)} = (1 - c)M + cR, \qquad \qquad J(x) = \frac{I(x) - c}{1-c}
$$ 
First note that the maximal matrix  $R$ is the same for $M$ and $N$, since the highest payoff to A in each column of $M_I$ is the same for all values of $I$.  Also, $J(x)$ will be continuous, nondecreasing, and concave down if $I(x)$ is, and $J(0) = 0$ since $I(0) = c$. 
Additionally, $J(x)$ will reach 1 at the same $x$ value that $I(x)$ does.

Then for any $x$, $N_{J(x)} = (1 - J(x)) N + J(x)R$ \\
= $(1 - \frac{I(x) - c}{1-c}) ((1 - c)M + cR) + \frac{I(x) - c}{1-c} R$ \\
= $(1 - c)M - (I(x) - c)M + cR - cR \frac{I(x) - c}{1-c} + R \frac{I(x) - c}{1-c}$ \\
= $(1 - I(x))M + cR + (I(x) - c)R$ \\
= $(1 - I(x))M + I(x)R$\\
Which is  $\mix{x}$ by definition.
 $\blacksquare $ \\
Therefore it is sufficient to only consider oracle functions with $I(0) = 0$.
For the remainder of this paper, we assume without loss of generality that all oracle functions are continuous, nondecreasing, (weakly) concave down, and satisfy $I(0) = 0$.

\section{Main Results} \label{mainresults}

We first show the conditions under which an equilibrium exists for an Oracle Game, followed by the conditions for a rational strategy to become dominated when enough information has been purchased. We then find the conditions and properties of the transitions occuring in these games as strategies become dominated or undominated.

\begin{Theorem}
\label{Equilibrium Conditions}
If $I(x)$ is differentiable at $c$ in the interior of its domain, then $\{s_a, s_b, c\}$ is an equilibrium of $\gi$ if and only if\\
1. $\{s_a, s_b\}$ is an equilibrium of $\mix{c}$ \\
2. $V(s_b) \cdot I'(c) = 1$ 
\end{Theorem}

\textbf{Proof:} 
Condition 1 holds if and only if player A or B have no incentive to change $s_a$ or $s_b$ respectively. 
If we express player A's payoff as $E_a(s_a, s_b, I(x)) - x$, then it suffices to find a global maximum of this function on its domain.  Taking the derivative with respect to $x$ and setting equal to zero yields 
$$
\frac{\partial E_a}{\partial I} \frac{dI}{dx} - 1 = 0
$$ 
assuming that $s_b$ is constant.  Since $V =  \partial E_a/ \partial I$ by definition, this is equivalent to condition 2, and shows that it yields a local maximum.   $V \geq 0$ and $I$ is (weakly) concave down imply that $E_a(s_a, s_b, I(x)) - x$ is also concave down with respect to $x$, so any local maximum must be a global maximum.  $\blacksquare $ \\

\begin{Lemma}
\label{Endpoints Lemma}
1.  If $\{s_a, s_b\}$ is an equilibrium of $\mix{0}$ and $\lim_{x \rightarrow 0^+} I'(x) \leq \frac{1}{V(s_b)}$ then $\{s_a, s_b, 0\}$ will be an equilibrium of $\gi$. \\
2.  If $\{s_a, s_b\}$ is an equilibrium of $\mix{x_1}$ and $\lim_{x \rightarrow x_1^-} I'(x) \geq \frac{1}{V(s_b)}$ then $\{s_a, s_b, x_1\}$ will be an equilibrium of $\gi$.
\end{Lemma}

\textbf{Proof:}  Although $I(x)$ will not be differentiable at the endpoints, (i.e. at $0$ and $x_1$ since player A cannot choose values of $x < 0$ and gains no benefit beyond $I(x) = 1$), we only need to look at the one sided limit in these cases.  If $\lim_{x \rightarrow 0^+} I'(x) \leq \frac{1}{V(s_b)}$, then player A will gain less benefit from increasing the oracle payment than the increase in cost, and has no incentive to do so.  Similarly, if $\lim_{x \rightarrow x_1^-} I'(x) \geq \frac{1}{V(s_b)}$, then player A will lose more benefit from decreasing the oracle payment than the reduction in cost, (and can gain no more benefit from increasing the cost, since $I$ is capped at 1), so has no incentive to change it. $\blacksquare $ \\

We also note that, even if $I'(x)$ has discontinuities, condition 2 of Theorem \ref{Equilibrium Conditions} can be modified to say that $c$ must equal the supremum over all points with $V(s_b) \cdot I'(x) \leq 1$.

For the remainder of the paper, we assume $M$ is a payoff matrix such that $\mix{x}$ has a unique Nash equilibrium for each $x$, except possibly at nodes.  Then we can define $s_a(x)$ and $s_b(x)$ as the strategies $s_a$ and $s_b$ in the unique equilibrium of $\mix{x}$ for all $x$ except at nodes.  If $x$ is a node and all equilibria at that node have the same $s_a$ or $s_b$, then $s_a(x)$ or $s_b(x)$ are defined as the appropriate strategy, while if $s_a$ or $s_b$ vary across equilibria, then the corresponding function is undefined at that node (in most games we consider, $s_a(x)$ will be defined at nodes and $s_b(x)$ will not).

We also add the assumption that $I$ is strictly increasing and strictly concave down.

 \begin{Proposition}
 \label{Strategy Loss}
If strategy $s$ for player B is not dominated in $M$ (weakly or strongly), but is dominated in $\mix{w}$ for some $w$, then it is strictly dominated for all $\mix{x}$ with $x > w$.  That is, a strategy which becomes dominated as $x$ increases remains dominated with further increase of $x$.
 \end{Proposition}

 \textbf{Proof:} 
First consider the case when $s$ becomes dominated by some pure strategy $p$.  Let $r_j$ be the payoffs to B for strategy $j$ in the matrix $R$ (A's best strategies when the oracle responds).  Let $b_{i,j}$ be the entry in $M$ in the $i$th row and $j$th column for B's payoff, then the entry in the $i$th row and $j$th column of  $\mix{x}$ will be 
$$
c_{i,j,x} = (1-I(x))b_{i,j} + I(x)r_j.
$$  
Define $m_{i,j} = r_j - b_{i,j}$, then $c_{i,j,x} = b_{i,j} + I(x) m_{i,j}$.  That is, the entries in the matrix will scale linearly with $I$, going from $b_{i,j}$ when $I = 0$ and reaching $r_j$ when $I = 1$.  Thus if strategy $s$ is not dominated by $p$ when $x = 0$ (and $I(0) = 0$), this means there must be a row $k$ such that $b_{k,j} \geq b_{k,p}$.  But if it is then dominated by $p$ for some nonzero payment $w$, this means that $b_{k,s} + m_{k,s} I(w) \leq b_{k,p} + m_{k,p} I(w)$.  Together these imply that $m_{k,p} > m_{k,s}$ and thus $r_p > r_s$. 

Then for any row $i$, $s$ dominated by $p$ at $w$ implies  
$b_{i,s} + m_{i,s} I(w) \leq b_{i,p} + m_{i,p} I(w)$.

Case 1: $b_{i,s} > b_{i,p}$.  Using the same argument as above, we have $m_{k,p} > m_{k,s}$.  Then each of $b_{i,s} + m_{i,s} I(w)$ and $b_{i,p} + m_{i,p} \cdot I(w)$ can be viewed as linear function dependent on $I$, with slope $m$.  Then if line $p$ has a greater slope and is above line $s$ at $I(w)$, then for any $x > w$, it will also be greater at $I(x)$ since $I$ is an increasing function.

Case 2: $b_{i,s} \leq b_{i,p}$  Going back to $c_{i,j,x} = (1-I(x))b_{i,j} + I(x)r_j$, that is, for every $x$, the elements $c_{i,j,x}$ are weighted averages of $b_{i,j}$ and $r_j$.  Then since $b_{i,s} \leq b_{i,p}$ and $r_s < r_0$, then 
$$
(1-I(x))b_{i,s} + I(x)r_s <  (1-I(x))b_{i,p} + I(x)r_p
$$
for all values of $x$.  

For any mixed strategy $p$, we can consider a hypothetical pure strategy whose payoffs are the weighted averages of the payoffs of its components (proportional to the probability that they are played).  If we let $r_p$ be the weighted average of the $r_j$ weighted by the frequencies of strategy $j$ in $p$ (instead of simply the best payoff to A in the hypothetical strategy), then the same analysis as above shows that if strategy $s$ becomes dominated by $p$ at some $w$, it will stay dominated by $p$ for all $x > w$.
 $\blacksquare $

Note that if a strategy starts out dominated at $x = 0$, Proposition \ref{Strategy Loss} does not apply. However, if such a strategy becomes undominated at some $x>0$, and then be redominated by another strategy at a greater $x$, Proposition \ref{Strategy Loss} would apply, and that strategy would remain dominated.  Thus each strategy corresponds to at most two nodes, and if $\gi$ has finitely many strategies implies, it has finitely many nodes.\\

\begin{Proposition}
 \label{Solution forms for s_a}
In each interval between two nodes, and for each strategy $i$ in the support of $s_a(x)$, there exist $a_i,b_i,c_i \in \mathbb{R}$ such that the probabilities of playing strategy $i$ in $s_a(x)$ can be expressed as $\frac{a_i+b_iI}{c_i(1-I)}$ for all $x$ in the interval.
 \end{Proposition}

 \textbf{Proof:} 
Let M be the payoff matrix.  Recall that the cross section matrix $\mix{x}$ has payoffs to B, $(1-I)b_{i,j} + I r_j$ where $b_{i,j}$ are the payoffs to B in M, and $r_j$ are the payoffs to B corresponding A's best response in column j.
Suppose $s_a(x) = (A_1, A_2, ... A_m)$ is a mixed strategy of Player A.  Then for each $j$, player B's expected value when playing strategy $j$ is 
$$
E_j = \sum_{i = 1}^{n} A_i [(1-I)b_{i,j} + Ir_j]
$$
If we fix a particular interval between two nodes, then the set of B's pure strategies which are undominated is constant on that interval.  Let $n$ be the number of undominated pure strategies for B in that interval.  Note that the condition that $\mix{x}$ has unique equilibria between nodes implies that A also has $n$ undominated pure strategies.  Fix $k$ as the index of one of B's undominated pure strategies.  Then $s_a(x)$ can be expressed as the solutions to the simultaneous equations $E_k = E_j$ for all $j \neq$ 1, and $\sum_{i = 1}^{n} A_i = 1$. 

Using the last condition, we obtain
$$
A_m = 1 - \sum_{i = 1}^{n-1} A_i
$$ 
which when substituted into the $E_j$ gives 
$$
E_j = 
\sum_{i = 1}^{n-1} A_i [(1-I)b_{i,j} + I r_j] +
(1 - \sum_{i = 1}^{n-1} A_i )[(1-I)b_{n,j} + Ir_j]
$$
$$
= (1-I)b_{n,j} + Ir_j +  \sum_{i = 1}^{n-1} A_i (1-I)(b_{i,j} - b_{n,j})
$$
Since every instance of $A_i$ is multiplied by $1-I$, we define $u_i = (1-I)A_i$ to get
$$
E_j = (1-I)b_{n,j} + Ir_j +  \sum_{i = 1}^{n-1} u_i (b_{i,j} - b_{n,j})
$$
Then we have $n - 1$ simultaneous equations with $n - 1$ variables $u_i$, with the coefficients on all $u_i$ in $\mathbb{R}$, and the constant terms have I with degree at most 1.  It follows that the solutions must be of the form $u_i = a + bI$ for some $a_i,b_i \in \mathbb{R}$.  Thus all $A_i$ are of the form $\frac{a_i+b_iI}{(1-I)}$ for some $a_i,b_i$ in $\mathbb{R}$.  This suffices to prove the proposition.  Additionally, if we have all $b_{i,j} \in \mathbb{Q}$, then $a_i,b_i \in \mathbb{Q}$, and by using the least common denominator of $a_i$ and $b_i$ we can express this as 
$$
A_i = \frac{a_i+b_iI}{c_i(1-I)}
$$ 
with $a_i,b_i,c_i \in \mathbb{Z}$ 
$\blacksquare $ \\

Note that this implies that $s_a(x)$ is continuous between nodes.  Additionally, the support of $s_a(x)$ must be constant between nodes because the support of $s_b(x)$ is. \\

\begin{Proposition}
 \label{s_b is constant}
The equilibrium strategy of Player B
$s_b(x)$ is piecewise constant with respect to $x$, with discontinuities only at the nodes.
 \end{Proposition}

\noindent
\textbf{Proof:} 
Suppose $\{s_a(c), s_b(c)\}$ is an equilibrium of $\mix{c}$ for some particular $c \in \mathbb{R}$.  Thus $s_b(c)$ causes player A to be indifferent among all strategies included in $s_a(c)$.  But since $s_a(c)$ is player A's strategy when the oracle does not respond, her indifference does not depend on I.  So $s_b(c)$ causes A to be indifferent on the support of $s_a(x)$ in $\mix{x}$ for all $x$.  Additionally $s_a(c)$ causes player B to be indifferent on the support of $s_b(c)$ in $\mix{c}$.

Let $d$ be any value such that there are no nodes between $c$ and $d$.  This means a strategy is dominated for B in $\mix{c}$ if and only if it is dominated in $\mix{d}$.  $s_b(c)$ is part of an equilibrium in $\mix{c}$, so none of the strategies in its support are dominated.  Thus they are also undominated in $\mix{d}$, so there must be some $s_a'$ which causes player B to be indifferent on the support of $s_b(c)$ in $\mix{d}$.  Proposition \ref{Solution forms for s_a} implies that the support of $s_a'$ is the same as the support of $s_a(c)$ since the formulas that define each strategy's probability are nonzero between nodes.  Then B must play a strategy that causes A to be indifferent on all strategies in this support.  $s_b(c)$ accomplishes this, thus ($s_a', s_b(c)$) is an equilibrium in $\mix{d}$.  And by assumption the equilibrium is unique at each point, so $s_b(c) = s_b(d)$
 $\blacksquare $ \\

Since we have expressed the equilibrium strategies $s_a(x)$ and $s_b(x)$ as functions of $x$, we can also express the expected payoff of player A as 
$$
E_a(x) = E_r(x) \cdot I(x) + E_n(x) \cdot (1 - I(x)) - x.
$$
where $E_r(x)$ is player A's expected payoff when $s_a(x)$ and $s_b(x)$ are played with the payoff matrix $R$ (the oracle responds) and $E_n(x)$ is A's expected payoff when $s_a(x)$ and $s_b(x)$ are played with the payoff matrix $M$ (the oracle does not respond).  Then the value of information $V = \frac{ \partial E_a }{ \partial I}$, which we showed earlier depends only on $s_b$, can also be expressed as a function of $x$:  
$$
V(x) = E_r(s_b(x)) - E_n(s_b(x)).
$$ 
It the follows immediately that $V(x)$ is piecewise constant with discontinuities only at the nodes, which comes from its direct dependence on $s_b(x)$. \\

 If we consider a simplified construction where a player has the binary option to purchase information or not for a fixed cost $c$, this corresponds to a stepwise oracle function $I(x)$.  But by proposition \ref{OracleEquivalence}, this is equivalent to an Oracle Game with a linear oracle function with slope $1/c$.  This, together with $V(x)$ piecewise constant and Theorem \ref{Equilibrium Conditions}, means equilibria will only occur at nodes (except when $c = 1/V$, in which case there are infinitely many equilibria)

The following Lemma is a stronger version of Proposition \ref{Strategy Loss} for strictly competetive games, as it eliminates the possibility of strategies  dominated at $x = 0$ which become undominated for $x>0$.  Thus, the only nodes that can occur are ones corresponding to strategies becoming dominated.

\begin{Lemma}
\label{Never Undominated}
If $G$ is strictly competitive, then any strategy for player B which is dominated in $M$ will be dominated in $\mix{x}$ for all x.
\end{Lemma}

\textbf{Proof:} 
Suppose that strategy $d$ dominates strategy $k$ for player B in $M$. Let $r_j$ be B's payoff in column $j$ of the maximal matrix R.  Since G is strictly competitive, this will also correspond to the lowest payoff for $B$ in column $j$ of M.  That is, $r_j \leq b_{i,j}$ for all $i, j$.  Then $r_t = b_{i,t}$ for some $i$.  Then $r_k \leq b_{i,k}$ and $k$ dominated by $d$ implies $b_{i,k} \leq b_{i,t}$.  And thus $r_k \leq r_t$.  Now for any $x$, let $c_{i,j,x}$ be the $i,j$ th entry for B in $\mix{x}$.  From the definition of $\mix{x}$, we get $c_{i,j,x} = (1-I(x))b_{i,j} + I(x)(r_j)$.  Then for any $i$, both $b_{i,k} \leq b_{i,t}$ and $r_k \leq r_t$ implies that $c_{i,k,x} \leq c_{i,t,x}$.  Thus $k$ is dominated by $d$ in $\mix{x}$.

Note that if $k$ is dominated by a mixed strategy, this argument extends in the same way as in Proposition \ref{Strategy Loss}. 
$\blacksquare $ \\

Note also that strict dominance in $M$ will imply strict dominance in $\mix{x}$. \\

\begin{Proposition}
 \label{Decreasing V}
If G is a strictly competetive game, then V(x) is nonincreasing with respect to x.
 \end{Proposition}

 \textbf{Proof:} 
Since $V(x) = E_r(x) - E_n(x)$, it is sufficient to show that $E_r(x)$ is nonincreasing and $E_n(x)$ is nondecreasing.  Let $E'_r$ be player B's payoff when the oracle responds, and $E'_n$ be his payoff when the oracle does not respond.  $G$ strictly competetive implies that $E_n$ is nondecreasing if and only if $E'_n$ is nonincreasing.  Since $R$ is made from entries in $G$, it is also a strictly competetive game matrix, so $E_r$ is nonincreasing if and only if $E'_r$ is nondecreasing.  So it suffices to show these properties for $E'_r$ and $E'_n$.  Note that these are both locally constant with discontinuities only at nodes since they are based on $s_b$.  Let $x'$ be any node.  Lemma \ref{Never Undominated} implies that this node occurs when a strategy becomes dominated, so let $j$ be one such strategy, and let $s$ be the (possibly mixed) strategy that dominates it at $x'$.  Then from the proof for Proposition \ref{Strategy Loss} we have $r_j < r_s$.  This means that $j$ is dominated in the matrix $R$.  Then, when player B shifts some of his mixed strategy probability from strategy $j$ to strategy $s$ as $x$ passes $x'$, $E'_r$ will increase.  Thus at every node, $E'_r$ must increase, and since it is constant on intervals between nodes, we conclude that $E'_r$ is nondecreasing.

Since $E'_n$ is also constant except at the nodes, the only place where it could possibly increase would be at a node.  Suppose, for the sake of contradiction, that $E'_n$ increases at the node $x'$.  Let $s_1$ be player B's strategy before the node, and $s_2$ be player B's strategy after the node.   For any value of $I$, we have
$$
E_b(s_b) = I \cdot E'_r(s_b) + (1-I)E'_n(s_b).
$$
Then $E_n$ increasing at $x'$ implies $E'_n(s_2) > E'_n(s_1)$.  We also showed above that $E'_r(s_2) > E'_r(s_1)$.  Both of these mean that for any value of I, $E_b(s_2) > E_b(s_1)$, which means $s_2$ will always yield a higher payoff to player B than $s_1$, assuming player A chooses $s_a$ optimally, regardless of how often the oracle responds.  This contradicts the assumption that $s_1$ was part of an equilibrium before the node, since player B could achieve a higher payoff by switching to $s_2$ immediately.  Therefore $E'_n$ must be nonincreasing.
 $\blacksquare $ \\

\begin{Theorem}
 \label{Unique Nash}
If $\mix{x}$ has a unique equilibrium for each $x$ except at nodes, $I$ is strictly concave down, and V(x) is nonincreasing, then $\gi$ will have a unique equilibrium.
 \end{Theorem}

\textbf{Proof:} 
Since we assume that each $\mix{x}$ has a unique equilibrium $\{s_a , s_b\}$, this covers condition 1 of Theorem \ref{Equilibrium Conditions}, except at the nodes.  It suffices to show that there is exactly one value of $x$ that satisfies condition 2 of Theorem \ref{Equilibrium Conditions}, and if it is a node then there is only one $\{s_a , s_b\}$ that still meets condition 1.

We can express the change in the expected value with increasing payment $x$ for Player A 
\begin{equation}
\frac{ \partial E_a }{ \partial x} = 
\frac{ \partial E_a }{ \partial I} \cdot \frac{dI}{dx} - 1 = VI' - 1
\label{e-dvaluedx}
\end{equation}
Since $I'$ is everywhere continuous and $V$ is continuous except at nodes, the expression will be continuous except at nodes.  $I$ nondecreasing and strictly concave down imply $I$ is strictly increasing, and thus $I' \geq 0$, and is strictly decreasing.  We've previously shown $V \geq 0$, and is nonincreasing.  These together imply $\frac{ \partial E_a }{ \partial x}$ is strictly decreasing.

For any $x$, $\{s_a(x),s_b(x)\}$ satisfy the first condition in Theorem \ref{Equilibrium Conditions}, by definition.  We now demonstrate that there is exactly one value of $x$ that satisfies either Lemma \ref{Endpoints Lemma} or the second condition of Theorem \ref{Equilibrium Conditions}:

\medskip
\noindent
Case 1: $\frac{ \partial E_a }{ \partial x}(0) < 0$.

This satisfies Lemma \ref{Endpoints Lemma}, and $\frac{ \partial E_a }{ \partial x}$ strictly decreasing means it is negative for all $x$, so there are no values of $x$ that satisfy $VI' = 1$, which means that $\{s_a(0),s_b(0),0\}$ will be the unique equilibrium of $\gi$.

\medskip
\noindent
Case 2: There exists a $c$ such that $\frac{ \partial E_a }{ \partial x}(c) = 0$.

Then $\{s_a(c), s_b(c), c\}$ satisfies Theorem \ref{Equilibrium Conditions}, and is an equilibrium for $\gi$.
Since $\frac{ \partial E_a }{ \partial x}$ is strictly decreasing, then for any $x < c$ we get $\frac{ \partial E_a }{ \partial x} > 0$, and for any $x > c$, $\frac{ \partial E_a }{ \partial x} < 0$ so this equilibrium is unique.

\medskip
\noindent
Case 3: $\frac{ \partial E_a }{ \partial x}(x)$ changes from positive to negative discontinuously at node $z$.

Let $s_1 = s_b(c_1)$ where $c_1$ is any value in the region immediately below $z$, and let $s_2 = s_b(c_2)$ where $c_2$ is any value in the region immediately above $z$. Using 
Eq.~\ref{e-dvaluedx}, we have $V(s_1) < \frac{1}{I'}$, and $V(s_2) > \frac{1}{I'}$, so there exists $p \in (0,1)$ such that 
$$
pV(s_1) + (1-p)V(s_2) = \frac{1}{I'}
$$  
Let $\beta = ps_1 + (1-p)s_2$.  Since $V$ is linear, this implies $V(\beta) = \frac{1}{I'}$, and thus $V(\beta)I' = 1$.  Since some strategy gets dominated at $z$, we have $supp(s_2) \subset supp(s_1)$, and thus $supp(\beta) = supp(s_1)$.

Now let $\alpha =  \lim_{x \rightarrow z^-} s_a(x)$, the strategy approached by Player A as $x$ approaches node $z$ from below.  Since $s_a(x)$ in this region make B indifferent on all strategies in $supp(s_1)$ in $\mix{x}$, then  $\alpha$ will also make B indifferent on all strategies in $supp(\beta)$ of $\mix{z}$ since this is preserved by the limit.  And B indifference to a strategy despite it being weakly dominated can only occur when one of A's strategies goes to probability 0 at $z$.  In particular, $supp(\alpha) = supp(s_a(c_2)) \subset supp(s_a(c_1))$.  Then, since $s_1$ makes A indifferent on all strategies in $supp(s_a(c_1))$,  and $s_2$ makes A indifferent on all strategies in $supp(s_a(c_2))$, then $\beta$ will make A indifferent on all strategies in $supp(\alpha)$.  Thus $\{\alpha,\beta \}$ is an equilibrium of $\mix{z}$.  Further, only linear combinations of $s_1$ and $s_2$ will make A indifferent on $supp(\alpha)$, and of those, only $\beta$ sets $VI' = 1$, so this equilibrium is unique.

Case 4: $\frac{ \partial E_a }{ \partial x} > 0$ for all $x$ up until $x_1$ such that $I(x_1) = 1$

\noindent
This satisfies Lemma \ref{Endpoints Lemma}.  Additionally, for any $x < x_1$ we have  $\frac{ \partial E_a }{ \partial x} > 0$ in which case A could profit from increasing $x$.  Thus $\{s_a(x_1),s_b(x_1),x_1\}$ will be the unique equilibrium of $\gi$, and there is no reason that Player A would ever pay more than $x_1$.  \\

Finally, we note that since
$V$ is continuous everywhere except at the nodes, positive, and weakly decreasing, when combined with $I'(x)$ strictly decreasing, implies that $\frac{ \partial E_a }{ \partial x}$ is continuous everywhere except at nodes, and strictly decreasing.  Thus exactly one of these cases must occur, depending on if and where $\frac{ \partial E_a }{ \partial x}$ changes from positive to negative.
$\blacksquare $ \\

\section{Harmful Information}

In most games we have considered so far, lower cost oracle functions (ones with greater $I(x)$) cause A's payoffs to increase when compared to more expensive ones.  However, this is not always the case.   We now consider a particular game matrix where cheaper information can be harmful to player A (in terms of decreasing his payoff in the equilibrium).  Let $G$ be the game given by the payoff matrix $M$:

\begin{table}[h]
\renewcommand{\arraystretch}{1.1}
\Large
\centering
  \begin{tabular}[h]{r|c|c|c|}
 \multicolumn{1}{c}{\diagbox[height=1.9em]{\normalsize 1}{\normalsize
    ~2}} & \multicolumn{1}{c}{$B_1$} & \multicolumn{1}{c}{$B_2$} 
        &\multicolumn{1}{c}{$B_3$} \\\cline{2-4}
    $A_1$   & $4,-1$ & $0,2$ & $0,0$  \\\cline{2-4}
    $A_2$ &  $0,2$ & $4,-1$ & $0,0$ \\ \cline{2-4}
  \end{tabular}
\end{table}

This essentially is a weighted matching pennies game where player B has the option to avoid playing altogether (opt out) by choosing strategy $B_3$.  In this classic normal form, the only Nash equilibrium is when A plays $(\frac{1}{2},\frac{1}{2})$ and B plays $(\frac{1}{2},\frac{1}{2},0)$, with expected values $E_A = 2$ and $E_B = \frac{1}{2}$. Player B will not choose strategy 3 because she can gain a nonzero amount of points by playing the mixed strategy.

If we consider the same strategies and payoffs but in a sequential game where B has to choose first, this is equivalent to giving A complete information about what B chooses (such as an oracle with the constant function $I(x) = 1$).  If we look at the tree this creates:

% tree layout
\tikzstyle{level 1}=[level distance=20mm, sibling distance = 40mm]
\tikzstyle{level 2}=[level distance=20mm, sibling distance = 25mm]
\tikzstyle{level 3}=[level distance=20mm, sibling distance = 13mm]
% node styles
\tikzstyle{player} = [circle, draw, inner sep=1.2, fill=black]
\tikzstyle{end} = [circle, draw, inner sep=1.2, fill=black ]

%\newcommand{\payoff}[4][below]{\node[#1] at (#2) {$(#3,#4)$};}
%already defined in an earlier tree

 \begin{figure}[h]
    \centering
\begin{tikzpicture}[grow=down]
      \node(0)[player]{} 
      child{node(1)[player] {} 
        child{node[player] {} edge from parent node [left]{$A_1$} 
        }
        child{node[player] {} edge from
          parent node [right] {$A_2$} 
        } 
        edge from parent node [above left] {$B_1$} 
      } 
      child{node(2)[player] {}
        child{node[player] {} edge from parent node [left] {$A_1$} 
        }
        child{node[end] {} edge from parent node [right] {$A_2$} 
        } 
        edge from parent node [right] {$B_2$} 
      }
    child{node(3)[player] {}
        child{node[player] {} edge from parent node [left] {$A_1$} 
        }
        child{node[end] {} edge from parent node [right] {$A_2$} 
        } 
        edge from parent node [above right] {$B_3$} 
      };

      %% information set

      %% specify players
      \node[above]at(0){$\mathbf{B}$}; 
      \node[right]at(2){$\mathbf{A}$};   
      \node[left] at (1) {$\mathbf{A}$};

      %% specify payoffs
 \payoff{1-1}4{-1}
 \payoff{1-2}02 
 \payoff{2-1}02
 \payoff{2-2}4{-1} 
\payoff{3-1}00
\payoff{3-2}00

    \end{tikzpicture}

    \caption{Harmful Information Extensive form game}
  \end{figure}
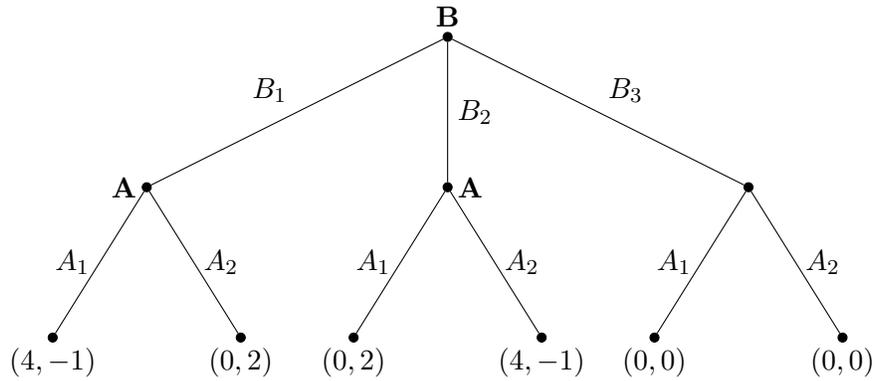

If B ever chooses $B_1$ or $B_2$, A will choose the best response and the payoff will be $(4,-1)$.  Knowing this, player B will only ever choose $B_3$, and both players will get a payoff of 0.  Note that this is worse for both players than the mixed strategy was.  That is, player A knowing what strategy player B has played is actually detrimental to both players.  If it were possible, player A would prefer not to have that information, or to be able to commit to ignoring the information and play a mixed strategy anyway, so as to incentivize player B to playing $B_1$ or $B_2$. The information however cannot be ``unseen". \\

How is this reflected in the Oracle Game $\gi$?  If A is given access to oracle $I(x)$, the payoff matrix $\mix{x}$ becomes:

\begin{table}[h]
\renewcommand{\arraystretch}{1.1}
\Large
\centering
  \begin{tabular}[h]{r|c|c|c|}
 \multicolumn{1}{c}{\diagbox[height=1.9em]{\normalsize 1}{\normalsize
    ~2}} & \multicolumn{1}{c}{$B_1$} & \multicolumn{1}{c}{$B_2$} 
        &\multicolumn{1}{c}{$B_3$} \\\cline{2-4}
    $A_1$   & $4,-1$ & $4I,2-3I$ & $0,0$  \\\cline{2-4}
    $A_2$ &  $4I,2-3I$ & $4,-1$ & $0,0$ \\ \cline{2-4}
  \end{tabular}
\end{table}

The equilibrium will occur in one of the following cases:\\

Case 1: If $I'(0) \leq \frac{1}{2}$ the oracle is too expensive to be worth paying, and the equilibrium is $\{(\frac{1}{2},\frac{1}{2}), (\frac{1}{2},\frac{1}{2},0), 0\}$ with expected values  $E_a = 2$ and $E_b = \frac{1}{2}$\\

Case 2 (interval): If $I'(0) \geq \frac{1}{2} \geq I'(x_\frac{1}{3})$ the equilibrium is $\{(\frac{1}{2},\frac{1}{2}), (\frac{1}{2},\frac{1}{2},0),y_\frac{1}{2}\}$, and neither player adjusts strategies due to the symmetry between strategy 1 and 2.  Then A's expected value is $E_a = 2 + 2I(y_\frac{1}{3}) - y_\frac{1}{2}$.  Note that since A is deliberately trying to maximize this, he's only paying the oracle when the function has steepness at least 2, so this payoff is greater* than his payoff in case 1 (*if the oracle is a straight line of slope $\frac{1}{2}$ it will be equal).  Player B's payoff is $E_B = \frac{1}{2} - \frac{3}{2} I(y_\frac{3}{2})$, which is worse than his payoff in Case 1, but still more than 0.\\

Case 3 (node): If $\frac{1}{2} \leq I'(x_\frac{1}{3})$ the equilibrium is $\{(\frac{1}{2},\frac{1}{2}),(\frac{1}{4I'},\frac{1}{4I'},\frac{4I'-2}{4I'}),x_\frac{1}{3}\}$.   When $I$ reaches $\frac{1}{3}$, B becomes indifferent between all three strategies, since her expected value from any of them is 0, so she would be willing to play any mixed strategy involving them.  But the only equilibrium is one where A is indifferent between $A_1$ and $A_2$ and also indifferent on increasing $x$ any further.  B's strategy in the equilibrium is the only one that satisfies both conditions.  In this equilibrium, Player B's expected value is $E_b = 0$, while Player A's is $E_a = \frac{4}{3I'} - x_\frac{1}{3}$.  Note that $E_a > 0$ because $I$ is concave down and must have slope at least $\frac{1}{2}$ in order to get to case 3, so $x_\frac{1}{3}$ will be small.  However $E_a$ is decreasing with respect to $I'(x_\frac{1}{3})$.  The reason for this is because A gets more payoff the more often B plays strategies 1 and 2, but as $I'(x_\frac{1}{3})$ increases, player B will opt out (strategy 3) more often in order to satisfy the condition $VI' = 1$.   As $I'(x_\frac{1}{3})$ approaches infinity, B's strategy will approach $(0,0,1)$, causing $E_a$ to approach 0. \\

%%%%%%%%%%%%%%%
\begin{figure}[t]%H
\center{
\includegraphics[scale = 0.28]{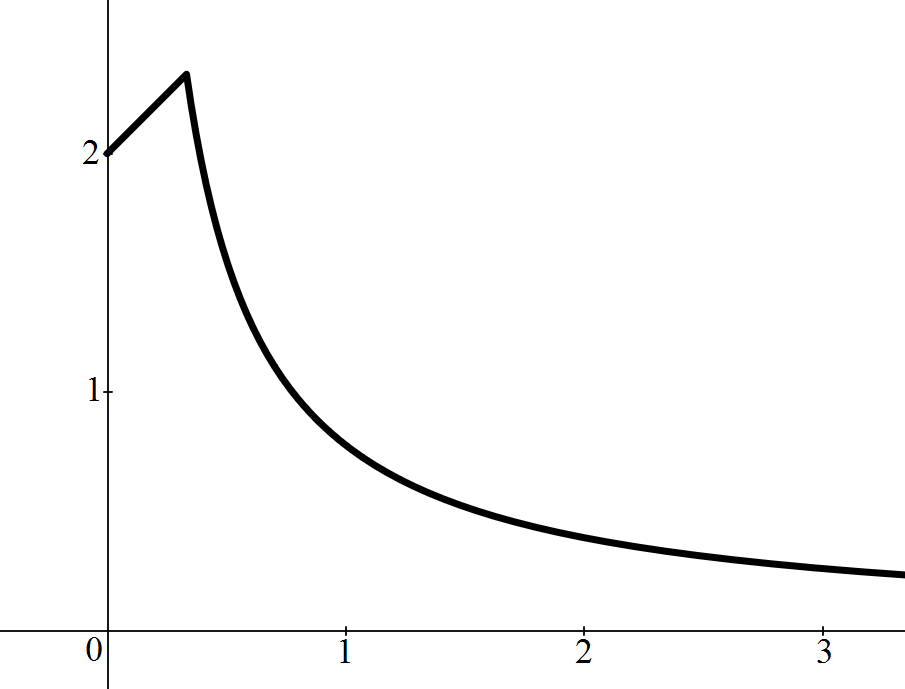}
\caption{Expected payoff $E_a$ as a function of $k$ for the oracle response function $I = \sqrt{kx}$.}
}
\label{HarmfulInfoEa}
\end{figure}
%%%%%%%%%%%%%%%

Thus, A will benefit the most at the boundary between case 2 and case 3.  
Figure 8 %\ref{HarmfulInfoEa} 
shows how her expected payoff changes as information becomes cheaper (as $k$ increases).  If the oracle is very expensive she will have to lose most of her benefit from the information to the oracle's cost.  But if information is too cheap then player B will be dissuaded from playing $B_1$ and $B_2$, and A will receive a lower payoff than if the oracle did not exist in the first place.  Similar to the sequential game, player A having cheap access to too much information decreases her expected value because player B will play the safer strategy to avoid getting exploited.  However because the Oracle Game provides a continuum of information to be purchased, our model demonstrates that having access to small amounts of information is beneficial to player A, while it is only when a certain threshold is reached that the information become harmful, by incentivizing player B to change strategies (as shown in Fig.~8).

\section{Multiple Equilibria} \label{multipleequilibria}

Although so far we have restricted our focus to games with one mixed equilibrium, games with multiple equilibria will tend to behave in a similar way: each equilibrium can be analyzed separately using the same techniques.
Consider the game defined by:

\begin{table}[H]
\renewcommand{\arraystretch}{1.1}
\Large
\centering
  \begin{tabular}[h]{r|c|c|c|c|}
 \multicolumn{1}{c}{\diagbox[height=1.9em]{\normalsize 1}{\normalsize
    ~2}} & \multicolumn{1}{c}{$B_1$} & \multicolumn{1}{c}{$B_2$} 
        &\multicolumn{1}{c}{$B_3$}  &\multicolumn{1}{c}{$B_4$} \\\cline{2-5}
    $A_1$   & $1,-1$ & $0,0$ & $-10,-10$ & $-10,-10$  \\\cline{2-5}
    $A_2$ &  $0,0$ & $2,-2$ & $-10,-10$ & $-10,-10$ \\ \cline{2-5}
    $A_3$ &  $-10,-10$ & $-10,-10$ & $2,-2$ & $0,0$ \\ \cline{2-5}
    $A_4$ &  $-10,-10$ & $-10,-10$ & $0,0$ & $3,-3$ \\ \cline{2-5}
  \end{tabular}
\end{table}

With no oracle, this game has three mixed strategy equilibria: one where they play their first two strategies, one where they play their last two strategies, and one where both players play all four strategies.  If an oracle is introduced with oracle function $I(x)$, there will still be three mixed strategy equilibria: two corresponding to the equilibria induced by the submatrix with either the first two strategies or the last two strategies of both players, and one involving both sets.  In this last case, some strategies might not be included, for instance if the oracle is cheap enough that $B_2$ or $B_4$ become dominated.  The two equilibria corresponding to smaller submatrices will be identical to the equilibrium in a game that was just that matrix with oracle function $I(x)$.    Thus, most of our results can be adapted and applied separately to each individual equilibrium in larger games.

\section{Conclusion}

%- include poster ideas

The Oracle Games defined here provide a method for investigating how players pay to acquire information, as well as how players respond to information about them being acquired. The probabilistic correction information in our model allows for detailed analysis, including the marginal increase in expected payoff which we call the value of information. We have shown that distinct transitions occur at nodes, where one of player B's strategies becomes dominated or undominated, and that these nodes are important in considering which strategies will be played and how much information should be purchased. 

The oracle is a stand-in for any process which might or might not succeed in providing noise-free information about a player's action. There is no fake information in our model - when the oracle does not return information, the player knows that it has failed. 
This approach could provide insight into competition between decision-makers, who are not  aware of each other's strategies but can invest time or resources to attempt to attain them at some cost (and risk of failure). 
% have applications involving industrial espionage, or any situation where 
In general, our model may be most useful whenever information is difficult to acquire, but is always reliable once acquired.  For example, a firm hiring spies to steal files from their competitors will have to pay regardless of whether they succeed in their operations or not, but if they succeed the files would be unlikely to contain false information.

Although the model by \citet{solan2004} is similar to ours, it gives different results about what sorts of equilibria occur.  They find that if sufficiently reliable information can be purchased cheaply enough, the player will purchase it and act on it as if it were completely true. They also show that the information cost affects the game's equilibria only insofar as it determines whether information is purchased or not; the actual amount of information purchased, if any, depends only on the payoffs in the original game.  
%This corresponds to the amount that causes the player's strategies to become dominated, and is related to our notion of nodes.  
In our model, a player will purchase more information as it becomes cheaper in a continuous way, until the point where a node is reached. And the shape and values taken by $I(x)$ can play a significant role.  
%Thus our mechanism of purchasing a randomly supplied perfect information is distinct from purchasing noisy information, and leads to different results even when attached to the same games.  

Oracle Games are fundamentally asymmetric because only one player has access to the oracle and its information.  There is no straightforward way to directly extend this to a symmetric system where both players have an oracle, since one player must commit to a decision before the oracle can know his action and provide it to the other player.  More complicated constructions could potentially resolve this, such as having both players bid payments for the oracle, and only the player with the higher bid gets access to the information.  Alternatively, an extended game could be played with both players having multiple actions, and each player could pay the oracle to learn information about the other player's earlier actions.  These modifications, or other similar ideas, could lead to symmetric games which would likely share many equilibrium features of our asymmetric approach.

Finally, using randomly supplied accurate information instead of noisy signals could be investigated in other games with information acquisition.  This could simplify the analysis by eliminating the need for players to condition their actions on uncertain beliefs, while still retaining the incentive to increase the amount of information available.

\section*{Acknowledgement}

We thank Glenn Young for insightful discussions and helpful comments. 
This work was supported in part by National Science Foundation Grant CMMI-1463482.

%\bibliography{OracleGamesBib}

\theendnotes

\end{document}